\documentclass[a4paper,12pt,reqno]{article}

\usepackage{graphicx,amssymb,amsmath,amsthm,mathrsfs,bbm}
\usepackage{bxcjkjatype}
\usepackage{xcolor, colortbl}
\usepackage{longtable}
\usepackage{natbib}
\usepackage[hidelinks]{hyperref}

\theoremstyle{plain}
\newtheorem{theorem}{Theorem}
\theoremstyle{definition}
\newtheorem{remark}[theorem]{Remark}
\numberwithin{equation}{section}

\newcommand{\abs}[1]{\lvert #1 \rvert}

\newcommand{\norm}[1]{\lVert #1 \rVert}

\newcommand{\bs}{\boldsymbol}
\newcommand{\mvert}{\, | \,}

\newcommand{\eps}{\varepsilon}

\newcommand{\RR}{\mathbb{R}}

\newcommand{\mcv}{\mathcal{V}}
\newcommand{\mcx}{\mathcal{X}}

\newcommand{\mfc}{\mathfrak{C}}
\newcommand{\mfk}{\mathfrak{K}}

\newcommand{\compo}{\mfc}
\newcommand{\kan}{\mfk}
\newcommand{\vein}{\mcv}

\newcommand{\one}{\mathbbm{1}}

\newcommand{\hbit}{\hspace*{1.5pt}}

\renewcommand{\thefootnote}{\fnsymbol{footnote}}
\title{Distance maps between Japanese kanji characters\\[0.5mm] based on hierarchical optimal transport}
\author{Dominic Schuhmacher\footnote{\href{mailto:schuhmacher@math.uni-goettingen.de}{schuhmacher@math.uni-goettingen.de}}\\[2mm]
Institute for Mathematical Stochastics\\
University of G\"ottingen}
\date{\today}

\begin{document}
\maketitle

\renewcommand{\thefootnote}{\arabic{footnote}}

\begin{abstract}
We introduce a general framework for assigning distances between kanji based on their dissimilarity. What we mean by this term may depend on the concrete application.
The only assumption we make is that the dissimilarity between two kanji is adequately expressed as a weighted mean of penalties obtained from matching nested structures of components in an optimal way. For the cost of matching, we suggest a number of modules that can be freely combined or replaced with other modules, including the relative unbalanced ink transport between registered components, the distance between the transformations required for registration, and the difference in prespecified labels.
  
We give a concrete example of a kanji distance function obtained in this way as a proof of concept. Based on this function, we produce 2D kanji maps by multidimensional scaling and a table of 100 randomly selected J\=oj\=o kanji with their 16 nearest neighbors.
  
Our kanji distance functions can be used to help Japanese learners from non-CJK backgrounds acquire kanji literacy. In addition, they may assist editors of kanji dictionaries in presenting their materials and may serve in text processing and optical character recognition systems for assessing the likelihood of errors.
\end{abstract}

\section{Introduction}

The Japanese language is widely considered to have one of the most complex writing systems in the world. In addition to two kana scripts of 48 basic characters each and their modifications via diacritics, an essential part are the kanji characters adopted and adapted from the traditional Chinese script. As a general rule, kanji are used for nouns, including proper names, for the front parts of adjectives and verbs (roughly their stems) and sometimes adverbs, whereas the kana scripts are used for other parts of speech, grammatical forms (including endings of adjectives and verbs), many foreign words and words where kanji are not available or do not seem appropriate for the intended audience.

A standardized list issued by the Japanese Ministry of Education currently specifies 2136 characters for habitual use (J\=oy\=o kanji).\footnote{See \url{https://www.bunka.go.jp/kokugo_nihongo/sisaku/joho/joho/kijun/naikaku/kanji/}} This list is intended as a guideline for written Japanese in everyday life, including newspapers and broadcasts. It is the required character set for official documents, and the basis of the kanji taught up to secondary level in the Japanese education system. Another list, issued by the Ministry of Justice, specifies an additional 863 characters that can be used for names in the Japanese family register (Jinmeiy\=o kanji).\footnote{See \url{https://www.moj.go.jp/MINJI/minji86.html}}

The set of kanji in actual everyday use differs substantially from the above lists. On the one hand, there is a smaller number of kanji that are rarely encountered (including some that are on the J\=oy\=o list because they are used in the Constitution of Japan). On the other hand, there is a larger number of kanji in common use, including terms from science, technology, art, and other specialized fields that are not on the J\=oy\=o list, raising the number of kanji known to many Japanese to over 3000. The highest two levels of the official Japan Kanji Proficiency Certification (Nihon Kanji Nōryoku Kentei, or short Kanken) require precise knowledge of around 3000 and 6000 characters, respectively. Comprehensive kanji dictionaries usually range between 5000 and well over 10000 entries. 
The free online KANJIDIC project\footnote{\url{http://www.edrdg.org/wiki/index.php/KANJIDIC_Project}} initiated by Jim Breen and maintained by the Electronic Dictionary Research and Development Group (EDRDG) provides information on the 13,108 kanji contained in the character sets of three major Japanese Industrial Standards for text processing (that form a subset of all CJK Unified Ideographs in Unicode). The largest kanji dictionary, the DaiKanWa Jiten, initially compiled by Tetsuji Morohashi and first published in 13 volumes between 1955 and 1966, contains about 50,000 characters (51,110 in its electronic version of 2018). 

For the foreign learner\footnote{Throughout this article, we use the term ``foreign learner'' for someone who has acquired functional knowledge of a first and possibly further writing systems unrelated to Chinese script. This assumes a rough lower age limit, but we do not mean to impose any other restrictions concerning age or ethnicity. Female pronouns are used for simplicity, without meaning to imply any particular gender.} of Japanese, the sheer amount of glyphs, estimated at around 1000, that need to be acquired to more or less understand an intermediate-level text may seem discouraging. However, the success story of spaced repetition systems (SRS), in combination with technological progress and the widespread availability of services over the internet, makes this process nowadays a rather painless experience for the ambitious student of Japanese. User statistics scraped from the major kanji learning website Wanikani\footnote{\url{https://www.wanikani.com/}} give a total of 10,143 users that have reached at least level 30 (roughly 1000 acquired kanji), whereas 2167 users are at level 60 (where they encounter the last of the 2074 kanji offered in Wanikani).\footnote{Data collected in 2022 by Wanikani user saraqael, see \url{https://community.wanikani.com/t/new-combined-user-statistics/58741}.} 

The full power of SRS has not been unleashed. For instance, many of them currently use fixed-time repetition intervals adapted only to the level of the item. 
Recent advances in artificial intelligence allow to optimize repetition times based on the user's capabilities, the complexity of the item learned, and the detailed history of item recall to make for an even more efficient and pleasant learning process.

A disadvantage of SRS is their inherent sequentiality. New items are presented in a particular order, which renders similarities and interrelations between subsequent kanji obvious (or at least easy to point out). But as the amount of kanji in the system increases, it becomes more and more difficult to robustly place newly presented kanji in the landscape of known kanji, i.e., to keep (consciously or unconsciously) track of similarities and differences to previous kanji to be able to distinguish them. 

As a particular example, consider the following four kanji from Wanikani: \begin{uCJK}粋\end{uCJK} (level~34), 
\begin{uCJK}酔\end{uCJK} (level 35), 
\begin{uCJK}枠\end{uCJK} (level 39) and
\begin{uCJK}砕\end{uCJK} (level 47). Presented next to each other, their similarity (the right-hand side consisting of characters \begin{uCJK}九\end{uCJK} for 9 and \begin{uCJK}十\end{uCJK} for 10) is easily noticed. The learner can quickly review them and sort out their different meanings and readings, probably with long-lasting effect. However, encountering the third and especially the fourth kanji in the SRS system (typically about two and four months, respectively, after first seeing their predecessors), it is rather unlikely that the learner will be aware of all the previous connections and there is a considerable risk that mix-ups occur when encountering any of the four kanji in later stages of the SRS or in the wild. Note also that the material meanings of the left-hand sides of these kanji (\begin{uCJK}米\end{uCJK} for rice, \begin{uCJK}酉\end{uCJK} for alcohol, \begin{uCJK}木\end{uCJK} for tree/wood and \begin{uCJK}石\end{uCJK} for stone) are likely to create a stronger anchor than the abstract numerals. But just once in a while, when seeing one of these kanji, 9 and 10 will stand out in the learner's mind, and the boundaries between the four kanji, both regarding their meanings and their (mostly very different) readings, will become blurred. 

For this and other cases, it is helpful if the learner can quickly look up similar kanji to find out what the other characters and their hidden connections are that interfere with her recalling of readings and meanings. It is also a good idea to search for similar kanji of an arbitrary learned kanji from time to time. Both strategies add struts and trusses to the shaky edifice of known kanji, allowing it to rise faster and higher into the sky. It has the pleasant side effect that the learner also encounters new kanji this way that can be attached to the edifice with little mental effort by contrasting the differing parts with similar learned kanji.

In the present article, I propose a mathematical framework for defining distances between kanji based on their dissimilarities. The last term is deliberately kept general and unspecific. Predicting to what extent somebody perceives two objects as similar or dissimilar is a complex problem discussed in various fields, including cognitive psychology, neuroscience, and pattern recognition in computer science. It can depend on multiple coarse and fine-grained individual factors, including prior knowledge of the objects, general education, socio-economic background, gender, and ethnicity. It may even vary considerably within each individual, depending on the mental and emotional state, previous associations, the current context and the mode of presentation.\footnote{It is well established among kanji learners that a relatively small change in font type can make all the difference between recognizing the kanji or not.} Addressing these aspects in sufficient detail is beyond this author's capabilities and will not be endeavored here. Instead, we aim in this paper for an abstract construction kit for building a kanji distance map, discuss various modules that can be used with it, and finally arrive at a concrete albeit somewhat ad-hoc distance map as proof of concept.

The utility of having adequate kanji distances available goes beyond helping the foreign learner build the trusses for her kanji edifice. Authors of textbooks and kanji dictionaries addressed to Non-Japanese obtain support in ordering and structuring the material they present; see \cite{Kandrac2021}. Text processing and optical character recognition systems may benefit in assessing the likelihood of potential mistakes based on whether other kanji fit the present context at close distance; see \cite{Nagata1998} and \begin{uCJK}阪本\end{uCJK}(Sakamoto) et al.~(2019). \nocite{ocr2019} 

We must clarify some terminology. Given a (typically large) set of kanji~$\mcx$, we refer to any symmetric function $d \colon \mcx \times \mcx \to \RR_+$ 
as \emph{distance map} if it satisfies the identification property that for any $\kan^{(1)}, \kan^{(2)} \in \mcx$
\begin{equation}
  d(\kan^{(1)}, \kan^{(2)}) = 0 \quad \text{if and only if} \quad \kan^{(1)} = \kan^{(2)}.
\end{equation}  
If also the triangle inequality holds, i.e., if
\begin{equation}
  d(\kan^{(1)}, \kan^{(2)}) \leq d(\kan^{(1)}, \kan^{(3)}) + d(\kan^{(3)}, \kan^{(2)}) \quad \text{for all $\kan^{(1)}, \kan^{(2)}, \kan^{(3)} \in \mcx$},
\end{equation}
the map $d$ is called a \emph{metric}. In both cases we refer to the values $d(\kan^{(1)}, \kan^{(2)})$ simply as distances. The term \emph{similarity} is mostly used in a loose sense in this paper. But since our distance maps are usually bounded by 1, the reader may very well think of $1-d(\kan^{(1)}, \kan^{(2)})$ as the similarity between $\kan^{(1)}$ and $\kan^{(2)}$ and of $1-d$ as a similarity map.

\subsection{Other sources of kanji distances or similarities}

The need for referencing and looking up similar kanji is well-recognized by various kanji learning tools. We have another look at Wanikani, which lists similar kanji for many of its characters. These similarities are often very accurate from the user's point of view, as they are presumably hand-picked. However, this comes with the downside that there are typically several clear omissions as well as biases towards only taking up similarities in certain components. Continuing with the four kanji discussed above, similar Wanikani kanji as of February 23, 2023, were:
\medskip

\noindent
\begin{uCJK}粋\end{uCJK}: \begin{uCJK}枠\end{uCJK}\\
\begin{uCJK}酔\end{uCJK}: \begin{uCJK}配 酸 酢 酎 酌\end{uCJK}\\
\begin{uCJK}枠\end{uCJK}: \begin{uCJK}粋\end{uCJK}\\
\begin{uCJK}砕\end{uCJK}: ---
\medskip

\noindent
At lower levels, the similar kanji seem more complete, as the list is slowly growing over time.

The only online kanji dictionary the author is aware of that offers a search for similar kanji is called Niai\footnote{\url{https://niai.mrahhal.net/similar}}. While the source code and the actual scores are available on GitHub\footnote{\url{https://github.com/mrahhal}}, the exact scoring method is unpublished. Judging from the credit section on GitHub, it is highly likely based on component matching using the public decomposition data in KRADFILE/RADKFILE\footnote{\url{http://www.edrdg.org/krad/kradinf.html}} initiated by Jim Breen and maintained by the EDRDG. The method may therefore be a particular case of the construction described below, just with a similarity rather than a distance target. Results seem to be quite reliable and fully automated. Still, it appears that similar shapes of not exactly matching components are not easily recognized and that good matches are often intermixed with several unclear similarities. For the four kanji above, we obtained as of February 23, 2023 (results restricted to J\=oy\=o kanji):
\medskip

\noindent
\begin{uCJK}粋\end{uCJK}: \begin{uCJK}酔 枠 砕 降 真\end{uCJK}\\
\begin{uCJK}酔\end{uCJK}: \begin{uCJK}粋 枠 砕 傘 焼\end{uCJK}\\
\begin{uCJK}枠\end{uCJK}: \begin{uCJK}枯 染 粋 酔 植\end{uCJK}\\
\begin{uCJK}砕\end{uCJK}: \begin{uCJK}括 活 砂 硬 碑\end{uCJK}\\
\medskip

There does not seem to be much previous research regarding a methodological treatment of kanji distances or similarities, at least not in the literature written in English.
A notable exception is \cite{YenckenBaldwin2006, YenckenBaldwin2008}. 
In the former paper, similarity experiments were conducted and analyzed that included 179 participants across 20 languages, comprising first-language speakers of Chinese and Japanese, second-language learners of Chinese and Japanese, and participants who had no previous experience in Chinese script (non-CJK group). Relatively simple similarity methods were proposed, one using the $\ell_1$-distance of rendered images and the other a bag-of-radicals cosine similarity based on KRADFILE/RADKFILE. Except for the participants from the non-CJK group, the bag-of-radicals method showed good overall accordance (in terms of rank correlation)
with the similarity judgment obtained from the mean or median raters in the experiment.

In \cite{YenckenBaldwin2008}, the authors introduced more sophisticated methods. These include a bag-of-radicals-plus-shape distance map,
in which their earlier cosine similarity is set to 1 if the first numbers in the kanjis' SKIP codes (\citealp{Halpern2022}), which classify four basic kanji patterns, disagree. This distance map performed best in terms of overall rank correlation with the human similarity judgments. The authors further decomposed kanji into sequences and trees of strokes using the hierarchical data from \cite{ApelQuint2004}, which is nowadays maintained on the kanjiVG website.\footnote{\url{http://kanjivg.tagaini.net}} The resulting stroke edit and tree edit metrics did not perform particularly well in the correlation experiment but (much) better than the other quantities in identifying highly similar kanji pairs in two follow-up experiments.

A file with pairwise similarities stemming from the stroke edit metric is available from one of the authors's website.\footnote{\url{https://lars.yencken.org/datasets/kanji-confusion}} Extracting the ten most similar kanji for each of our four kanji above and transforming back from similarities to stroke edit distances,
we obtain Table~\ref{tab:k90neighbors_seditdist}.

\begin{table}
\begin{center}
\small
\begin{tabular} {|c||c|c|c|c|c|c|c|c|c|c|}
  \hline
  \raisebox{3pt}{\strut}\begin{uCJK}粋\end{uCJK} & 
  \begin{uCJK}断\end{uCJK} & \begin{uCJK}枠\end{uCJK} & \begin{uCJK}料\end{uCJK} & \begin{uCJK}粉\end{uCJK} & \begin{uCJK}析\end{uCJK} & \begin{uCJK}新\end{uCJK} & \begin{uCJK}数\end{uCJK} & \begin{uCJK}砕\end{uCJK} & \begin{uCJK}卒\end{uCJK} & \begin{uCJK}科\end{uCJK} \\
  & 0.1818 & 0.2000 & 0.2000 & 0.3000 & 0.3000 & 0.3077 & 0.3846 & 0.4000 & 0.4000 & 0.4000 \\
  \hline
  \raisebox{3pt}{\strut}\begin{uCJK}酔\end{uCJK} & \begin{uCJK}酢\end{uCJK} & \begin{uCJK}酌\end{uCJK} & \begin{uCJK}配\end{uCJK} & \begin{uCJK}酷\end{uCJK} & \begin{uCJK}酬\end{uCJK} & \begin{uCJK}酪\end{uCJK} & \begin{uCJK}卑\end{uCJK} & \begin{uCJK}枠\end{uCJK} & \begin{uCJK}車\end{uCJK} & \begin{uCJK}昇\end{uCJK} \\
  & 0.2500 & 0.2727 & 0.2727 & 0.2857 & 0.3077 & 0.3077 & 0.3636 & 0.3636 & 0.3636 & 0.3636 \\
  \hline
  \raisebox{3pt}{\strut}\begin{uCJK}枠\end{uCJK} & \begin{uCJK}析\end{uCJK} & \begin{uCJK}粋\end{uCJK} & \begin{uCJK}株\end{uCJK} & \begin{uCJK}柳\end{uCJK} & \begin{uCJK}砕\end{uCJK} & \begin{uCJK}断\end{uCJK} & \begin{uCJK}酔\end{uCJK} & \begin{uCJK}杉\end{uCJK} & \begin{uCJK}卒\end{uCJK} & \begin{uCJK}松\end{uCJK} \\
  & 0.1250 & 0.2000 & 0.3000 & 0.3333 & 0.3333 & 0.3636 & 0.3636 & 0.3750 & 0.3750 & 0.3750 \\
  \hline
  \raisebox{3pt}{\strut}\begin{uCJK}砕\end{uCJK} & \begin{uCJK}砲\end{uCJK} & \begin{uCJK}研\end{uCJK} & \begin{uCJK}枠\end{uCJK} & \begin{uCJK}許\end{uCJK} & \begin{uCJK}群\end{uCJK} & \begin{uCJK}粋\end{uCJK} & \begin{uCJK}破\end{uCJK} & \begin{uCJK}訴\end{uCJK} & \begin{uCJK}碑\end{uCJK} & \begin{uCJK}故\end{uCJK} \\
  & 0.3000 & 0.3333 & 0.3333 & 0.3636 & 0.3846 & 0.4000 & 0.4000 & 0.4167 & 0.4286 & 0.4444 \\
  \hline
\end{tabular}
\end{center}
\vspace*{-4mm}

\caption{\label{tab:k90neighbors_seditdist} The closest ten kanji of the four kanji and their stroke edit distances.}
\end{table}

\subsection{Contribution}

We introduce a general framework (Section~\ref{sec:toplevel_kdist}) for assigning distances between kanji that makes the rather minimalistic assumption that dissimilarity is adequately expressed by a weighted mean of penalties obtained from comparing nested structures of sets of strokes, which we call components, in an optimal way.

The penalties are obtained as a function of various old and new building blocks, such as a comparison of labels of components, their relative positions, sizes, or distortions, and how far ink has to be transported and how much of it has to be added or erased to transform one component into the other. These building blocks can be added and replaced according to our needs, such as the required specificity and level of sophistication or the computation time available.

We present a combination of useful building blocks in Section~\ref{sec:seclevel_wasser} and make first parameter choices to obtain a concrete distance map as a proof of concept in Section~\ref{sec:results}.
As a preview, we give the ten nearest neighbors and the corresponding distances (on a scale from 0 to 0.25) for our four kanji from above in Table~\ref{tab:k90neighbors_kdist}. Compared with Table~\ref{tab:k90neighbors_seditdist}, this table contains fewer kanji where the similarity with the target kanji is not so apparent, especially if we look at the right halves of the tables (nearest neighbors number 6--10).

\begin{table}
\begin{center}
\small
\begin{tabular} {|c||c|c|c|c|c|c|c|c|c|c|}
  \hline
  \raisebox{3pt}{\strut}\begin{uCJK}粋\end{uCJK} & \begin{uCJK}枠\end{uCJK} & \begin{uCJK}砕\end{uCJK} & \begin{uCJK}粒\end{uCJK} & \begin{uCJK}粉\end{uCJK} & \begin{uCJK}料\end{uCJK} & \begin{uCJK}粘\end{uCJK} & \begin{uCJK}枯\end{uCJK} & \begin{uCJK}粧\end{uCJK} &\begin{uCJK}粗\end{uCJK} &\begin{uCJK}辞\end{uCJK} \\
  & 0.0596 & 0.1210 & 0.1315 & 0.1323 & 0.1326 & 0.1345 & 0.1397 & 0.1397 & 0.1431 & 0.1437 \\
  \hline
  \raisebox{3pt}{\strut}\begin{uCJK}酔\end{uCJK} & \begin{uCJK}酢\end{uCJK} & \begin{uCJK}酎\end{uCJK} & \begin{uCJK}酌\end{uCJK} & \begin{uCJK}配\end{uCJK} & \begin{uCJK}酷\end{uCJK} & \begin{uCJK}鮮\end{uCJK} & \begin{uCJK}酬\end{uCJK} & \begin{uCJK}酪\end{uCJK} & \begin{uCJK}酸\end{uCJK} & \begin{uCJK}酵\end{uCJK} \\
  & 0.0594 & 0.1028 & 0.1056 & 0.1083 & 0.1187 & 0.1203 & 0.1234 & 0.1235 & 0.1268 & 0.1333 \\
  \hline
  \raisebox{3pt}{\strut}\begin{uCJK}枠\end{uCJK} & \begin{uCJK}粋\end{uCJK} & \begin{uCJK}枯\end{uCJK} & \begin{uCJK}枝\end{uCJK} & \begin{uCJK}砕\end{uCJK} & \begin{uCJK}染\end{uCJK} & \begin{uCJK}乗\end{uCJK} & \begin{uCJK}祥\end{uCJK} & \begin{uCJK}析\end{uCJK} & \begin{uCJK}朽\end{uCJK} & \begin{uCJK}杉\end{uCJK} \\
  & 0.0596 & 0.0875 & 0.1067 & 0.1109 & 0.1137 & 0.1312 & 0.1328 & 0.1400 & 0.1400 & 0.1407 \\
  \hline
  \raisebox{3pt}{\strut}\begin{uCJK}砕\end{uCJK} & \begin{uCJK}枠\end{uCJK} & \begin{uCJK}粋\end{uCJK} & \begin{uCJK}辞\end{uCJK} & \begin{uCJK}砂\end{uCJK} & \begin{uCJK}酔\end{uCJK} & \begin{uCJK}砲\end{uCJK} & \begin{uCJK}研\end{uCJK} & \begin{uCJK}計\end{uCJK} & \begin{uCJK}枯\end{uCJK} & \begin{uCJK}群\end{uCJK} \\
  & 0.1109 & 0.1210 & 0.1238 & 0.1483 & 0.1544 & 0.1549 & 0.1551 & 0.1552 & 0.1572 & 0.1583 \\
  \hline
\end{tabular}
\end{center}
\vspace*{-4mm}

\caption{\label{tab:k90neighbors_kdist} The closest ten kanji of the four kanji and their kanji distances.}
\end{table}

However, the focus of this article is not so much on the concrete distance map, which still has considerable room for optimization and should ideally be adapted to a specific task. 
Instead, it rests on presenting the overall construction as a multi-level matching of components, discussing what important targets for comparing components might be, and showcasing several modules that can be used for this comparison.

At the root of our method lies a modern structural approach that may be referred to as hierarchical optimal transport.
On the macroscopic level, we solve an optimal transport problem by matching components (which is itself hierarchical in nature due to the hierarchical component structure); but also when determining the cost between components we use optimal transport when computing the (relative unbalanced) Wasserstein distance between rendered images of components. This metric addresses the fundamental problem with the (relative) $\ell_1$ metric mentioned in \cite{YenckenBaldwin2006}. Rather than averaging the absolute pixel difference at all locations, it computes the average distance by which ink has to be transported to transform one image into the other. As a preview of a more detailed description, the reader may refer to the right panel of Figure~\ref{fig:brokensun}. Independently of the left panels, this can be seen as a difference plot of pixel images for the kanji \begin{uCJK}日\end{uCJK} (day, sun) and \begin{uCJK}臼\end{uCJK} (mortar). Although the kanji have been rescaled and aligned, the lines match hardly anywhere precisely (except for the small bits on the right towards the top and on the bottom towards the left, where the white positive and the black negative line cancel each other out). Consequently, the $\ell_1$ distance is relatively large, whereas the average distance of ink transport (indicated by the arrows) is quite small. 

Of course, the computational cost for a Wasserstein distance is much higher than for an $\ell_1$-distance. However, it is nowadays perfectly feasible to compute the transport in Figure~\ref{fig:brokensun} in well under 0.1\hbit s on an ordinary laptop (e.g., with the function \verb|unbalanced| from the \textsf{R} \nocite{rproject} package \textsf{transport}, see \citealp{transport}). It turns out that a much smaller resolution (e.g., $32 \times 32$ instead of $64 \times 64$ in Figure~\ref{fig:brokensun}) is sufficient. Also, if time is really of the essence, we can use a faster approximative computation method; see \cite{ChizatEtAl2018}. 

As a last contribution, we mention the new \textsf{R} package \textsf{kanjistat} (\citealp{kanjistat}), which offers a selection of computational and statistical tools for working with kanji. Among other things, it can import and convert several of the data sets mentioned in this paper and compute the various component distances described, as well as the overall kanji distance. Parts of the package are still a work in progress, and the whole package will be continuously improved in the future.

\section{Optimal transport between nested component structures} \label{sec:toplevel_kdist}

In this section, we focus on the abstract definition of our kanji distance map, leaving any more disputable design and parameter choices for later sections.

On a global level, we think of a kanji as built of (sometimes overlapping) components made up of strokes. A \emph{stroke} is an element of a product space that contains all the desirable information we want to represent in our kanji distance. Choosing $U=[0,1]^2$ as our canvas for drawing kanji, this would usually comprise at least one path $f \colon [a,b] \to U$ or one path-connected subset $F \subset U$, either of which represents the stroke in a single pre-specified font type. Further information may include the stroke number $s$, a stroke type classification $\tau$, and possibly more, yielding, e.g., a tuple $(f,s,\tau)$ to represent the stroke. A \emph{component} is a finite set of strokes.\footnote{To lighten the notation, we refrain from explicitly adding further information in the definition of the component, such as a label or the relative position of the component in the kanji. Formally this information could be included at the stroke level.}

We then model a kanji as a finite sequence $\kan = (\kan_l)_{0 \leq l \leq L+1}$ of increasingly fine decompositions $\kan_l = \bigl\{\compo_{l,i} \mvert 1 \leq i \leq m_l \bigr\}$ into components, starting with all strokes in a single component at level $l=0$ and ending with each stroke in an individual component at level $l=L+1$.
Writing $[m] := \{1,\ldots,m\}$, we require more precisely:
\begin{enumerate}
  \item[(1)] For every $l \in [L+1]$ and every $i \in [m_l]$ there is a $j$ in $[m_{l-1}]$ such that $\compo_{l,i} \subset \compo_{l-1,j}$ (\emph{nesting property across levels}).
  \item[(2)] For every $l \in [L+1]$ and $i,i' \in [m_l]$ with $i \neq i'$ we have $\compo_{l,i} \setminus \compo_{l,i'} \neq \emptyset$\\ (\emph{components on the same level are not nested}).
  \item[(3)] $\kan_0 = \{ \compo_{0,1} \} = \bigl\{ \{S_1, \ldots, S_n \} \bigr\}$.
  \item[(4)] $\kan_{L+1} = \{ \compo_{L+1,1}, \ldots \compo_{L+1,n} \} = \bigl\{ \{S_1\}, \ldots, \{S_n\} \bigr\}$.
\end{enumerate}
Properties (1) and (4) imply a \emph{covering property}, i.e., $\bigcup_{i=1}^{m_l} \compo_{l,i} = \{S_1, \ldots, S_n \}$ for every~$l$. While this union will typically be disjoint, our only intra-level requirement is (2). We refer to $\compo_{0,1}$ as the \emph{root} and to any of the $\compo_{L+1,i}$ as \emph{leaves} even though the inclusion relation does not necessarily give rise to a tree structure. 

The index $L$ denotes the last level of interest for our kanji distance. We are usually not interested in the decomposition into individual leaves at level $L+1$, as it does not contain more information than level $L$ (nor even level $0$). However, we could always add a copy of this decomposition as a new second-last level if we want to include the leaves anyway. Also, we can omit any level $l \in [L]$ to get a simpler hierarchy of decompositions. Neither step destroys the above properties. 

We refer to the set $K = \{(l,i) \mvert 1 \leq l \leq L, 1 \leq i \leq m_l\}$ of remaining indices as the \emph{essential index set} of the kanji. By a \emph{vein}, we mean any set $V = \{(l,i_l) \mvert 0 \leq l \leq L\} \subset K$ whose components form a maximal chain of inclusions from root to one-before-leaf, i.e., $\compo_{0,1} \supset \compo_{1,i_1} \supset \ldots \supset \compo_{L,i_L}$.

\begin{figure}
  \begin{center}
  \hspace*{-1mm}\includegraphics[width=.25\linewidth]{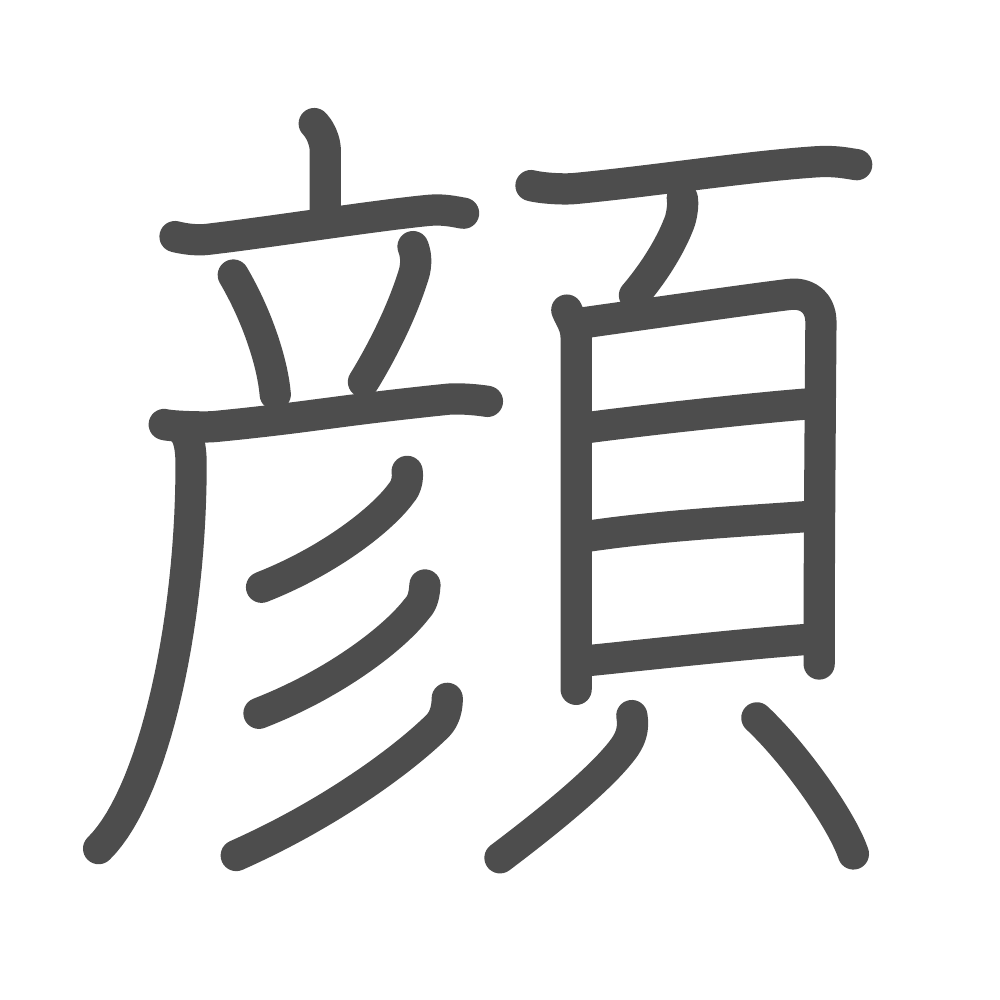}\hspace*{0mm}\includegraphics[width=.25\linewidth]{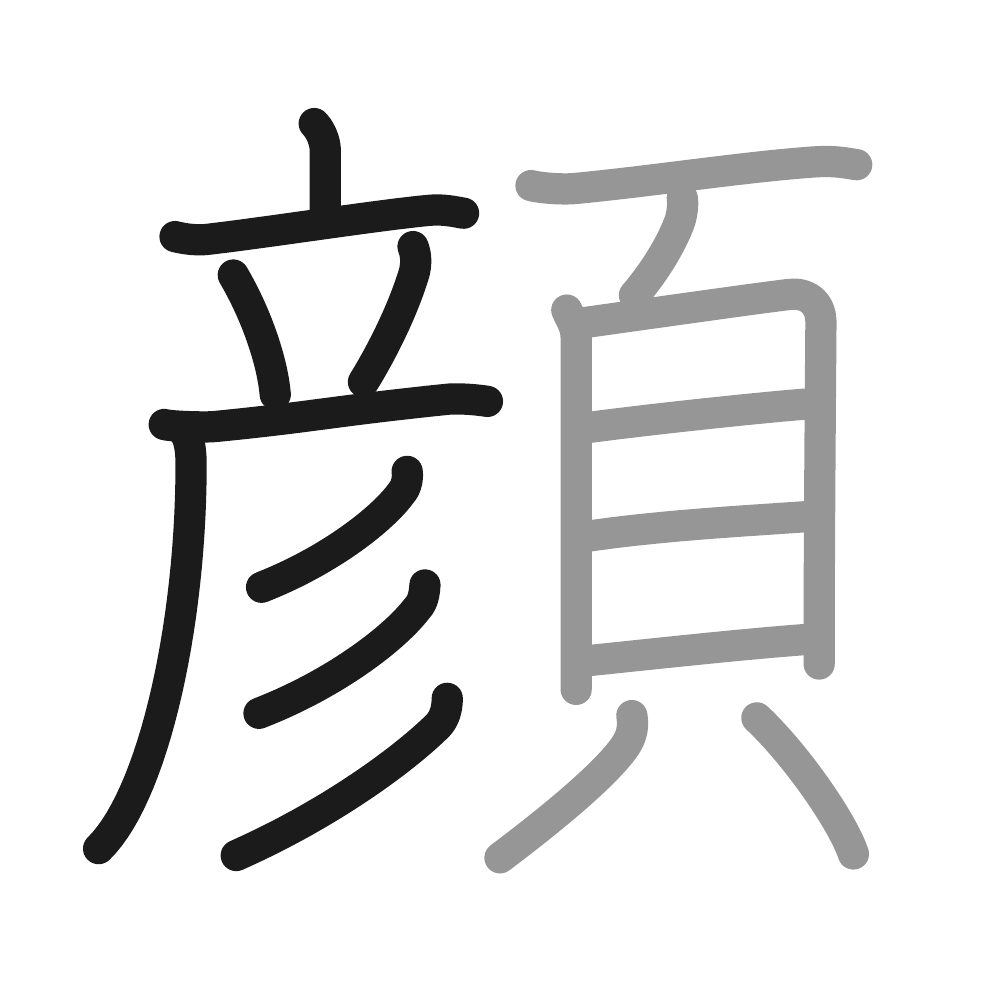}\hspace*{0mm}
  \includegraphics[width=.25\linewidth]{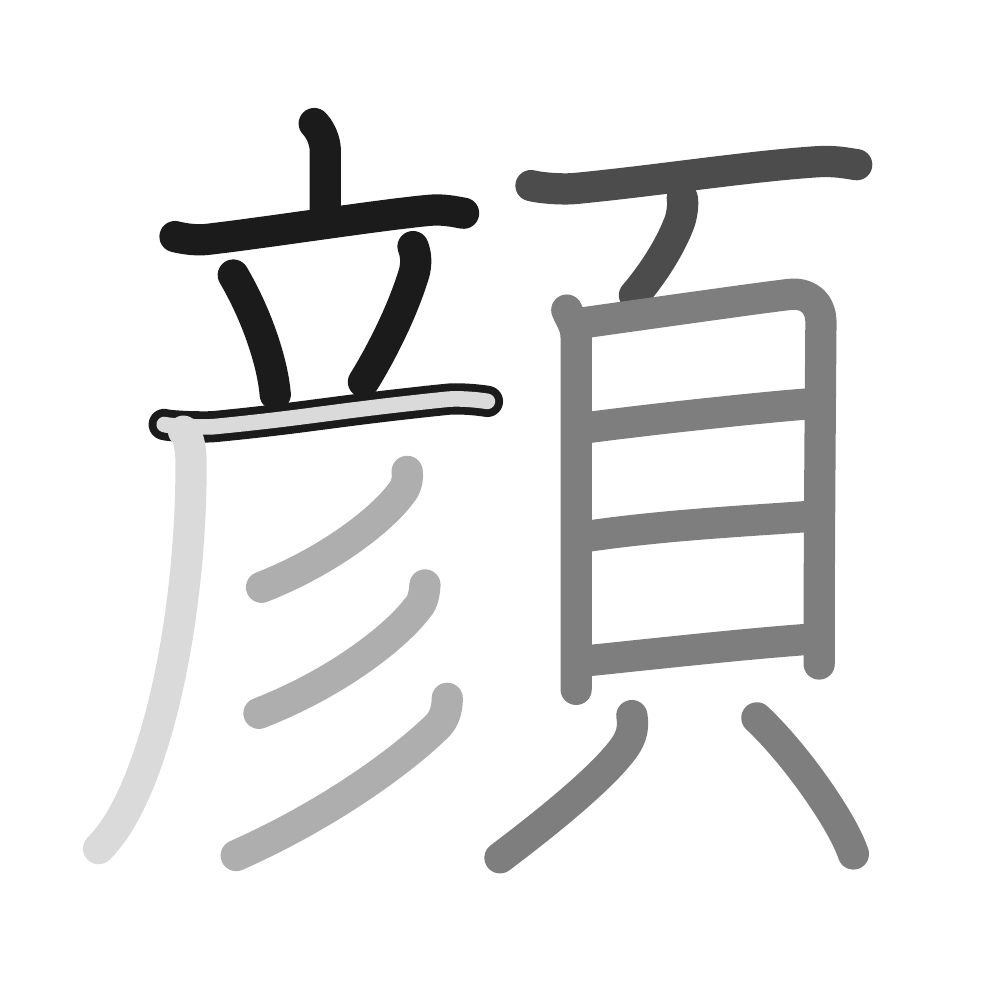}\hspace*{0mm}
  \includegraphics[width=.25\linewidth]{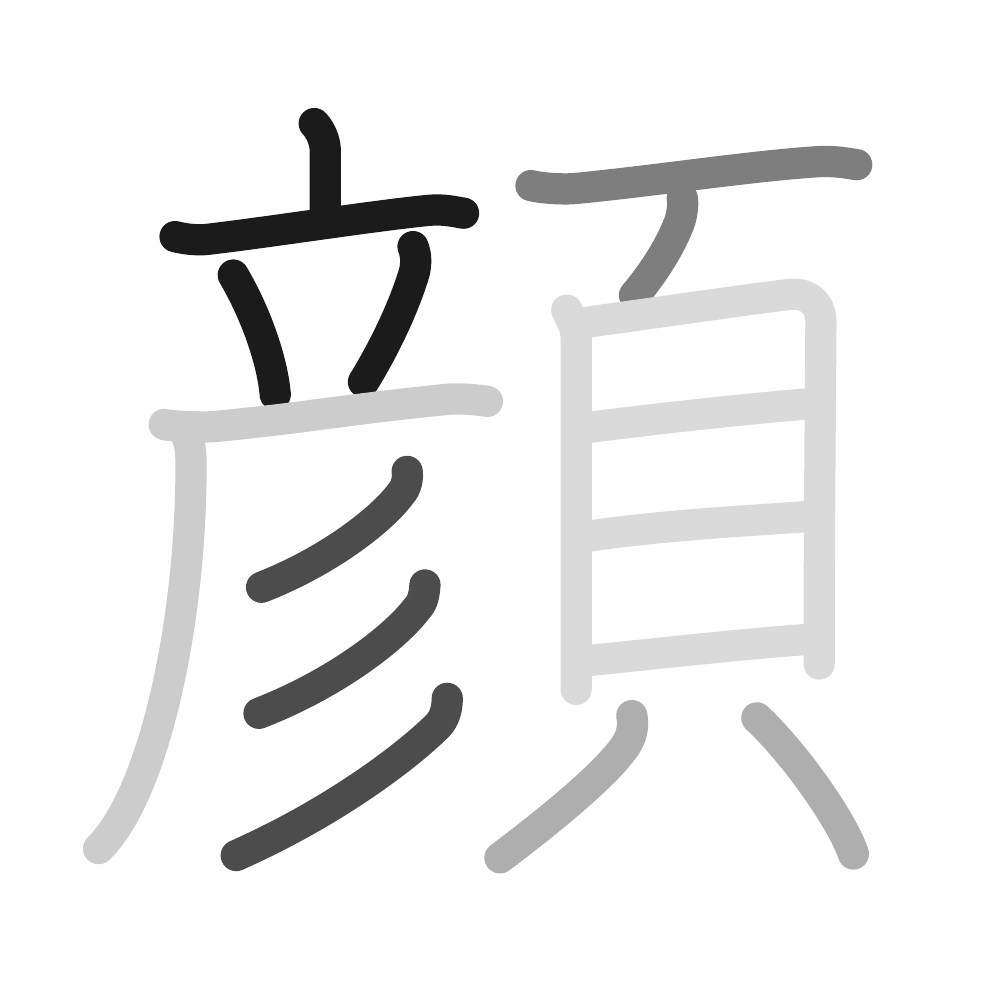}\hspace*{-1mm}
  \end{center}
  \vspace*{-10mm}
  
  \caption{\label{fig:kaodecomp}Decompositions of the kanji \begin{uCJK}顔\end{uCJK} at levels 0, 1, 2, 3 resulting in 1, 2, 5, 6 components, respectively. Grayscales distinguishing the components are randomly re-assigned for each level. Note the overlap of components \begin{uCJK}立\end{uCJK} and \begin{uCJK}厂\end{uCJK} at level 2. Not shown is level $4 = L+1$, which is just the decomposition of the kanji into its 18 individual strokes.}
\end{figure}

Figure~\ref{fig:kaodecomp} shows the decomposition hierarchy for the kanji \begin{uCJK}顔\end{uCJK} based on its kanjiVG file. Like all the examples in this paper, it can be reproduced with standard commands in the \textsf{R} package \textsf{kanjistat}. The figure illustrates that the decompositions need not be disjoint: at level 2, the bottom stroke of \begin{uCJK}立\end{uCJK} is shared with the top stroke of \begin{uCJK}厂\end{uCJK}, which is later resolved at level~3 when the top part of \begin{uCJK}立\end{uCJK} forms a component alone. Assuming we enumerate components from top to bottom and then left to right, examples of veins are given by the sets $\{(0,1), (1,2), (2,5), (3,5)\}$ corresponding to the components \begin{uCJK}顔\end{uCJK}, \begin{uCJK}頁\end{uCJK}, \begin{uCJK}貝\end{uCJK}, \begin{uCJK}目\end{uCJK} and $\{(0,1), (1,1), (2,2), (3,2)\}$ corresponding to the components \begin{uCJK}顔\end{uCJK}, \begin{uCJK}彦\end{uCJK}, \begin{uCJK}厂\end{uCJK}, \begin{uCJK}厂\end{uCJK}.

Our proposed kanji distance map is defined by matching components of two kanji across all levels as well as possible without producing any nestings. The distance is then given as a weighted sum of component distances, adding a penalty with the weight of uncovered parts of the kanji.

We first introduce an appropriate structure $w = (w_{l,i})_{(l,i) \in K}$ of (non-negative) weights satisfying $\sum_{i=1}^{m_l} w_{l,i} = 1$ for any~$l$ and describing the relative importance of the various components within each level. Typically, $w_{l,i}$ is an increasing function of some size aspect of component $\compo_{l,i}$, such as the relative amount of ink needed to draw the component. 
Since we use the line information from kanjiVG at constant width in this paper, the amount of ink for component $\compo_{l,i}$ is adequately expressed by summing up the stroke lengths $\abs{S}$ for $S \in \compo_{l,i}$.

We also include a factor $(1-\eps)$ for each level beyond $l=1$, representing the ``trickling away'' of weight as we descend in the component hierarchy. This creates a certain incentive to match larger components rather than their individual parts. In particular, it penalizes matchings of subcomponents that are not arranged in the right way to be matched en bloc. 
Our weights are then
\begin{equation}  \label{eq:compoweights}
  w_{l,i} = \frac{\sum_{S \in \compo_{l,i}} \abs{S}}{\sum_{S \in \compo_{0,1}} \abs{S}} (1-\eps)^{\max\{0,\,l-1\}}.
\end{equation}

Next, a symmetric function $\mu \colon [0,1] \times [0,1] \to [0,1]$ is required for determining the single contribution weight $\mu(w,w')$ obtained from matching two components with given weights $w$ and $w'$. We require that $\mu(w,w') \geq \min\{w,w'\}$ and $\sum_{i=1}^m \sum_{i'=1}^{m'} \mu(v_i, v'_{i'}) \leq 1$ for any finite weight vectors $(v_i)$, $(v_{i'})$ with $\sum_{i=1}^m v_i = \sum_{i'=1}^n v_i' = 1$, which allows for $\mu(w,w') = \frac{w+w'}{2}$ as a maximal choice. The geometric or harmonic means form a good compromise since they put more weight on matches with different individual weights than the minimum while still going to zero if one of the weights goes to zero (as opposed to the arithmetic mean).

Finally, we denote by $\varrho$ a $[0,1]$-valued distance map between components. Appropriate constructions will be discussed in Section~\ref{sec:bottomlevel_compodist}. Among other things, our $\varrho(\compo, \compo')$ will usually incorporate the minimal ink transportation cost between standardized drawings of $\compo, \compo'$ and penalties for the translation, scaling, and distortion necessary to make these drawings as similar as possible.

We can now define the kanji distance map $d$ as follows. Let $\kan^{(1)}, \kan^{(2)}$ be two kanji with nested decompositions $\bigl( \{ \compo^{(1)}_{l,i} \mvert 1 \leq i \leq m_l \} )_{0 \leq l \leq L+1}$ and $\bigl( \{ \compo^{(2)}_{l',i'} \mvert 1 \leq i' \leq m'_{l'} \} )_{0 \leq l' \leq L'+1}$, respectively. Denote by $K_1$ and $K_2$ the corresponding essential index sets and by $\vein_1$ and $\vein_2$ the sets of veins of $\kan^{(1)}$ and $\kan^{(2)}$. Let furthermore $w^{(1)} = (w^{(1)}_{l,i})_{(l,i) \in K_1}$ and $w^{(2)} = (w^{(2)}_{l',i'})_{(l',i') \in K_2}$ be the two weight structures defined according to~\eqref{eq:compoweights}. Let $a \in (0,1]$ and define
\begin{equation} \label{eq:kanjidist}
\begin{split}
  d(\kan^{(1)}, \kan^{(2)}) = \min_{(e_{(l,i),\,(l',i')})} &\biggl( \sum_{(l,i) \in K_1} \sum_{(l',i') \in K_2} e_{(l,i),\,(l',i')} \, \mu(w^{(1)}_{l,i}, w^{(2)}_{l',i'}) \, \varrho \bigl( \compo^{(1)}_{l,i}, \compo^{(2)}_{l',i'} \bigr) \\
  &\hspace*{1mm} + a \biggl( 1 - \sum_{(l,i) \in K_1} \sum_{(l',i') \in K_2} e_{(l,i),\,(l',i')} \, \mu(w^{(1)}_{l,i}, w^{(2)}_{l',i'}) \biggr) \biggr) \in [0,a],
\end{split}
\end{equation}
where the minimum is taken over $(e_{(l,i),\,(l',i')}) \in \{0,1\}^{K_1 \times K_2}$ satisfying
\begin{equation*}
\begin{split}
  \sum_{(l,i) \in V_1} \sum_{(l',i') \in K_2}   e_{(l,i),(l',i')} &\leq 1 \quad \text{for all $V_1 \in \mcv_1$}, \\
  \sum_{(l,i) \in K_1} \sum_{(l',i') \in V_2}   e_{(l,i),(l',i')} &\leq 1 \quad \text{for all $V_2 \in \mcv_2$}.
\end{split}
\end{equation*}

Thus $e_{(l,i),(l',i')} = 1$ signifies that components $\compo^{(1)}_{l,i}$ and $\compo^{(2)}_{l',i'}$ are matched and make an additive contribution of $\varrho \bigl( \compo^{(1)}_{l,i}, \compo^{(2)}_{l',i'} \bigr)$ weighted by $\mu(w^{(1)}_{l,i}, w^{(2)}_{l',i'})$. Any weight missing for a total contribution of 1 after taking all matches into account contributes~$a$. The side constraints mean that 
we can only ever match a single component per vein (once). In particular, if $e_{(1,0),(l',i')} = 1$ for some $(l',i') \in K_2$, i.e., the whole first kanji at the top level is matched (as might be reasonable, e.g., for \begin{uCJK}徴\end{uCJK} and \begin{uCJK}懲\end{uCJK}), we must have $e_{(l,i),(l'',i'')} = 0$ for all $(l,i) \in K_1 \setminus \{(1,0)\}$ and all $(l'',i'') \in K_2$, meaning we cannot match any subcomponent of \begin{uCJK}徴\end{uCJK} with the \begin{uCJK}心\end{uCJK} component of \begin{uCJK}懲\end{uCJK} (nor with anything else).

\begin{figure}
  \begin{center}
  \hspace*{-1mm}\includegraphics[width=.17\linewidth]{kao0.pdf}\hspace*{0mm}\includegraphics[width=.17\linewidth]{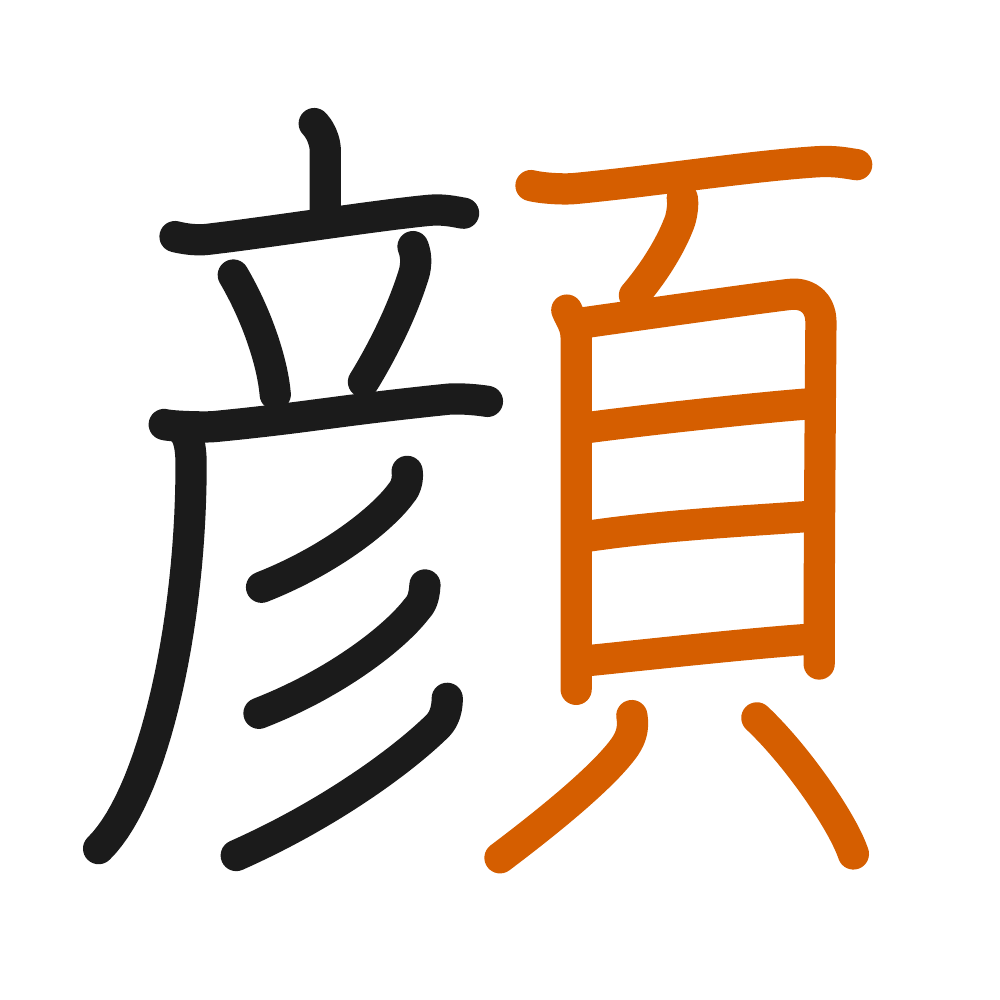}\hspace*{0mm}\includegraphics[width=.17\linewidth]{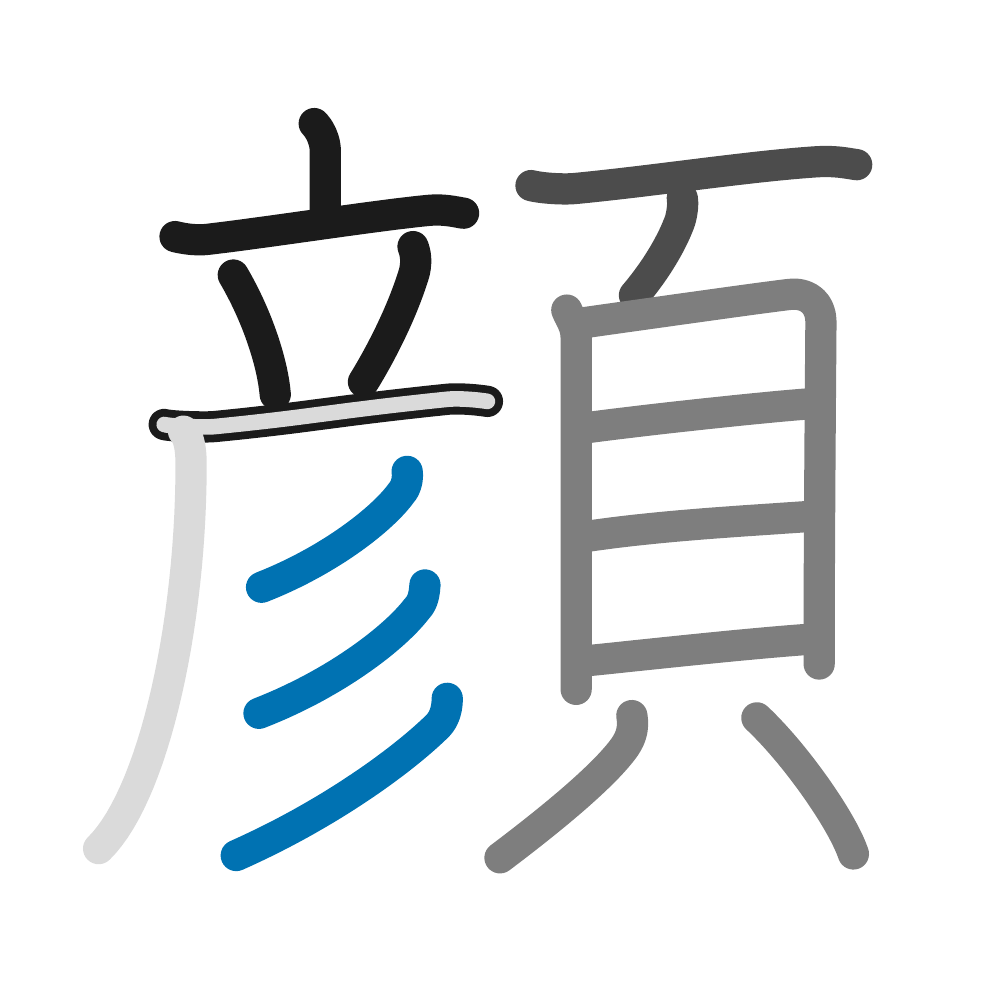}\hspace*{2mm}\includegraphics[height=6.5em]{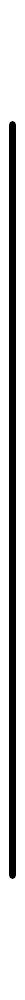}\hspace*{0mm}
  \includegraphics[width=.17\linewidth]{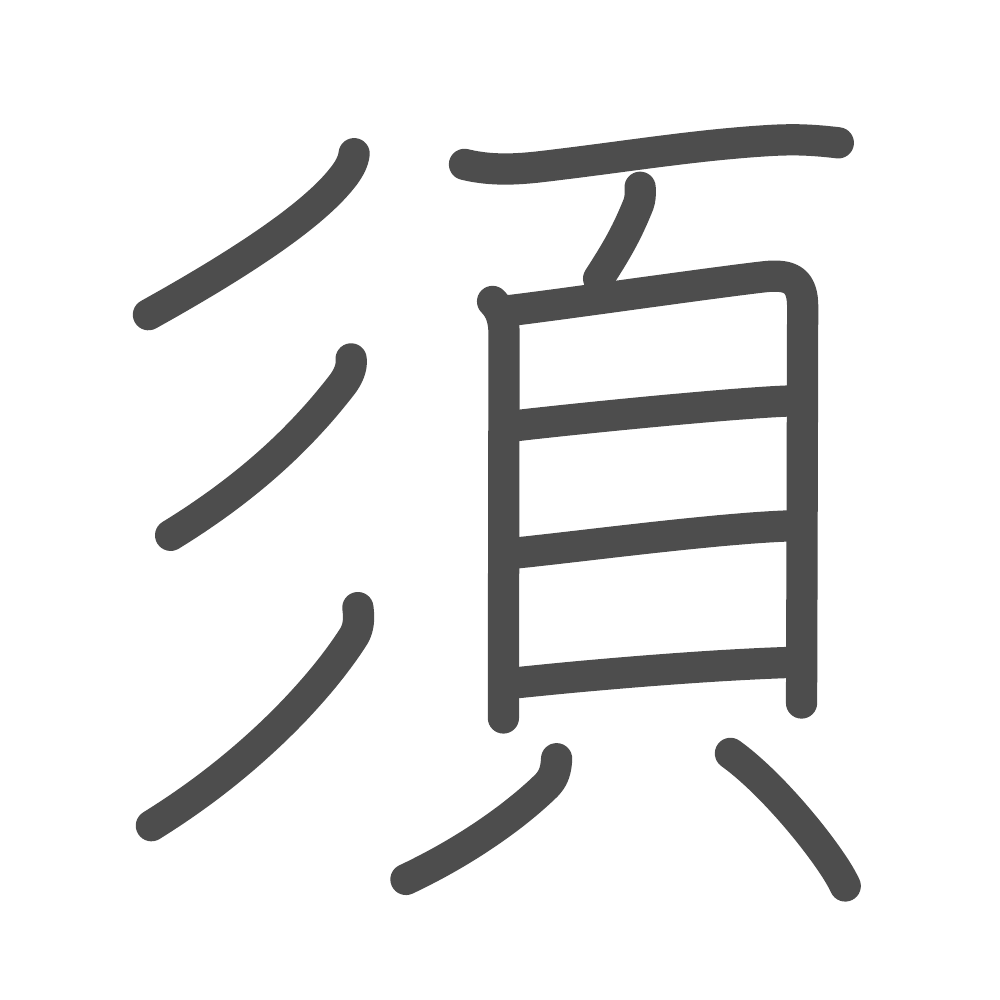}\hspace*{-1.5mm}\includegraphics[width=.17\linewidth]{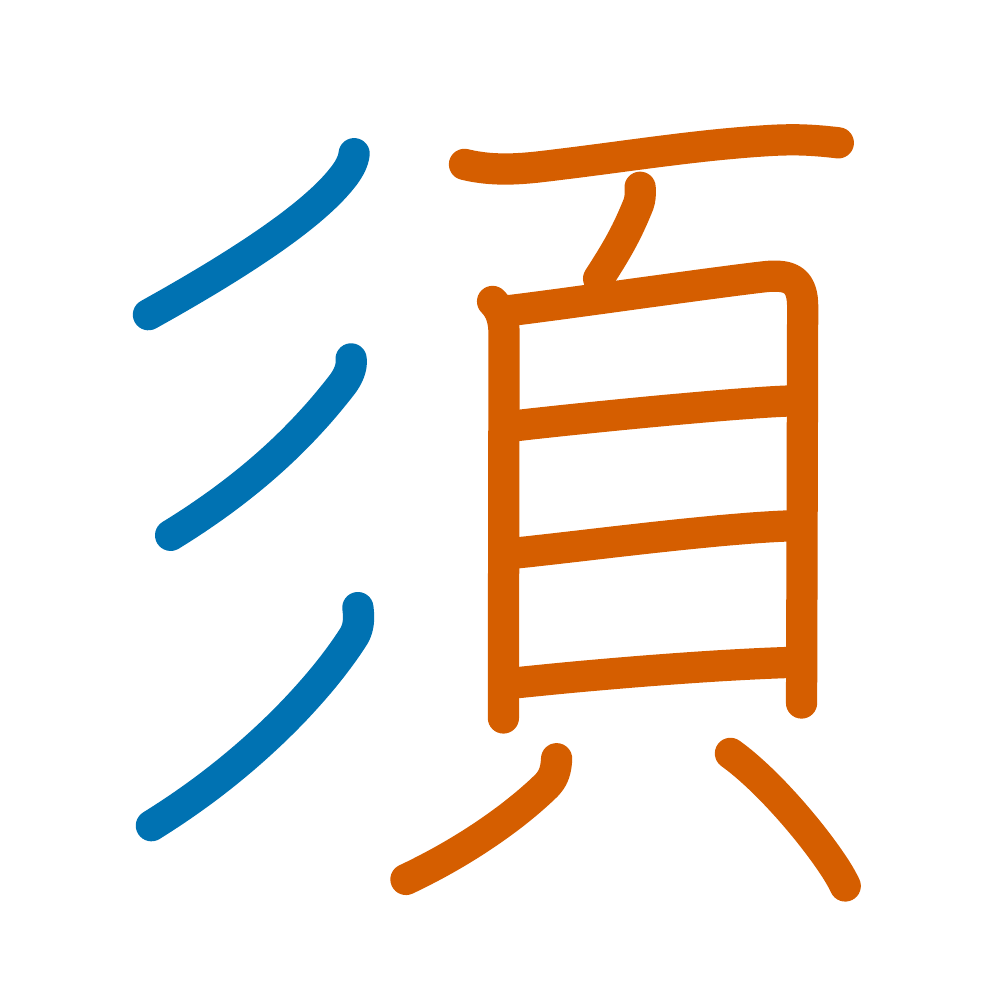}\hspace*{-1.5mm}\includegraphics[width=.17\linewidth]{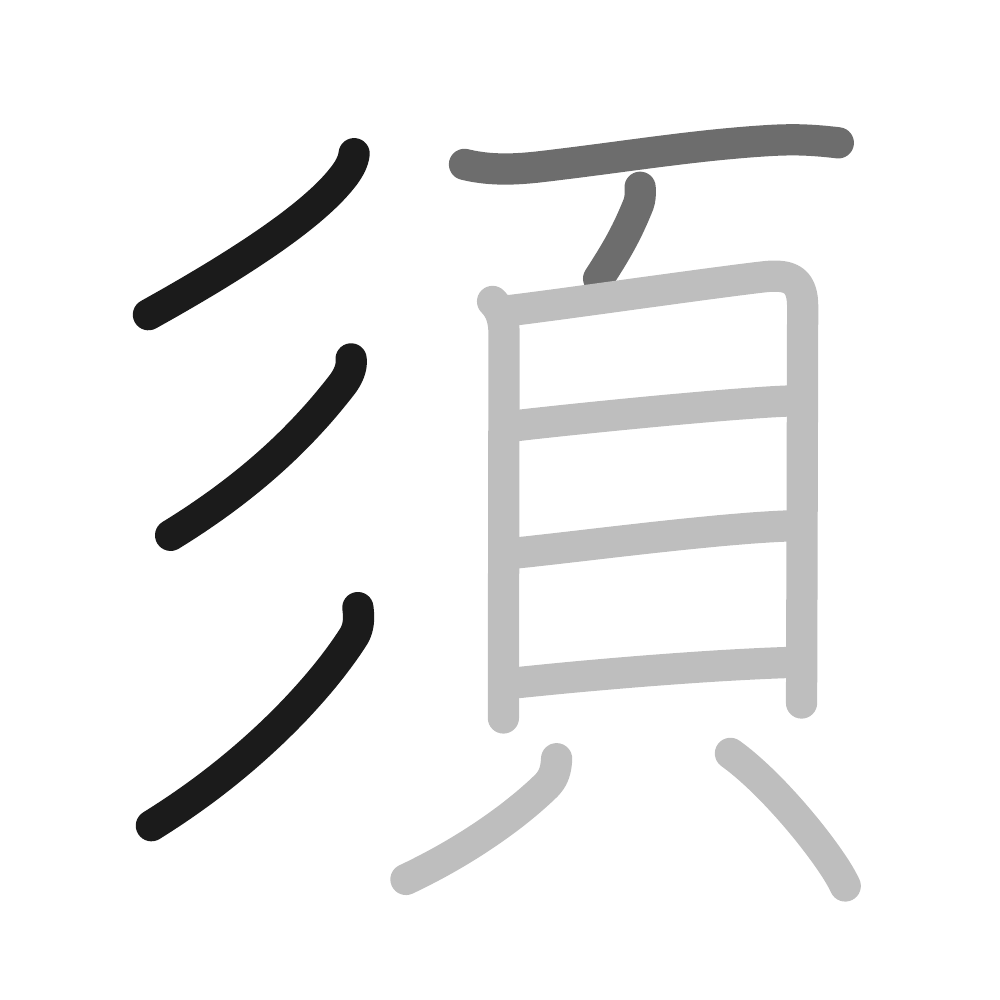}\hspace*{-1mm}
  \end{center}
  \vspace*{-9mm}
  
  \caption{\label{fig:kao_vs_shu}Decompositions of the kanji \begin{uCJK}顔\end{uCJK} and \begin{uCJK}須\end{uCJK} at levels 0, 1, 2 and their component matchings. The orange and blue colors indicate that components \begin{uCJK}頁\end{uCJK} and \begin{uCJK}彡\end{uCJK} are matched, i.e., $e_{(1,2), (2,2)} = 1$ and $e_{(2,3), (2,1)} = 1$. 
  The left-hand side kanji could still create a match for \begin{uCJK}立\end{uCJK} and \begin{uCJK}厂\end{uCJK} (or subcomponents thereof), but there is nothing left to match within the right-hand side kanji.}
\end{figure}

A slightly more complex example is shown in Figure~\ref{fig:kao_vs_shu} for the kanji \begin{uCJK}顔\end{uCJK} and \begin{uCJK}須\end{uCJK}. For the more concrete distance we use in Section~\ref{sec:results}, the components \begin{uCJK}頁\end{uCJK} and \begin{uCJK}彡\end{uCJK} are matched as indicated in the figure, the former at a very small cost, the latter at a somewhat higher cost.

The optimization problem \eqref{eq:kanjidist} is a binary linear program with linear side constraints, which can be solved with standard software. Since it is a small problem with usually fewer than 100--200 variables, it does not contribute substantially to the computation time. 

From a practical point of view, it may be desirable to prune the essential index sets (and hence the veins) further by omitting a number of indices at the end of certain inclusion chains according to some criterion, e.g., one might want to ignore components that have fewer than three strokes. This requires minor adaptations in the definition above. Either we allow more general essential index sets or incorporate the criterion into the distance map $\varrho$, e.g., by setting $\varrho(\compo,\compo') = a$ (or larger) if one of $\compo,\compo'$ is to be omitted.

\begin{remark}
  Without more specific assumptions, $d$ is not a metric. It is always a distance map, i.e., has the symmetry and identification properties 
  because $\varrho$ does.\footnote{Note that equality of components or kanji in the context of the identification property depends on what information we include (formally at the stroke level) when modeling the kanji. If we are not careful, it may happen that this equality is weaker than the equality of two kanji in a linguistic sense. E.g., if only stroke types but no stroke positions are included, we have $\text{\begin{uCJK}部\end{uCJK}}=\text{\begin{uCJK}陪\end{uCJK}}$ and $\text{\begin{uCJK}招\end{uCJK}}=\text{\begin{uCJK}拐\end{uCJK}}$ and $\text{\begin{uCJK}呆\end{uCJK}}=\text{\begin{uCJK}杏\end{uCJK}}$. The first two can be resolved by adding labels for the components, but the last one needs positional information. In the remainder of this article, we will always use enough information to identify components (and kanji) in the linguistic sense.} 
However, the triangle inequality requires more. Since our goal is to quantify (dis)similarity and since it is generally accepted that human similarity judgment and recognition tasks are often non-metric in nature, see e.g.\ \cite{Tversky1977}, \cite{LaubEtAl2006}, \cite{ScheirerEtAl2014} or \cite{NachshonEtAl2022}, the triangle inequality is not necessarily something we strive for.\\
On the other hand, it may be interesting to have a metric available when illustrating the kanji landscape (see Section~\ref{sec:results}) or for further mathematical study of the kanji space. Assuming that $\varrho$ is a metric and $\mu$ is the minimum, we \emph{conjecture} that \eqref{eq:kanjidist} defines a metric. If we refrain from matching across the hierarchy levels by fixing $l=L$ and $l'=L$, it can be \emph{shown} that we obtain a metric. The same remains true if we allow multi-matches of components according to split-up weights rather than performing a one-to-one matching according to minimum weights. 
 This would then be a similar overall construction as in~\cite{YurochkinEtAl2019}.
\end{remark}

\section{Comparing components} \label{sec:bottomlevel_compodist}

While the previous section provides an overall strategy for measuring distances between kanji by breaking down the problem to a multilevel comparison of components in terms of $\varrho$, the task of setting up a suitable function $\varrho$ remains. For the largest part, the components are no simpler than actual kanji. However, we may now concentrate entirely on the global features of the components --- their overall looks, shapes, areas, and positions --- trusting that the multilevel construction for $d$ from Section~\ref{sec:toplevel_kdist} takes care of any similarity or distance based on their substructures.

\subsection{General considerations}

As mentioned in the introduction, the author cannot offer a definitive answer on how different a single person or a larger group of people consider two kanji or two components, among other things, because of a lack of a good data foundation and any own psycholinguistic expertise. Nevertheless, I would like to go through a number of aspects and examples to make the reader aware of how multi-facetted the question is and sort out what aspects our function $\varrho$ defined below does and does not (but maybe should and definitely could) take into account.

As a starting point, consider the four components \begin{uCJK}牛\end{uCJK} (cow), \begin{uCJK}午\end{uCJK} (noon), \begin{uCJK}干\end{uCJK} (dry), \begin{uCJK}千\end{uCJK} (thousand). The reader may agree that at least any two direct neighbors in this series look quite similar. Ignoring the structural point of view (they all have at least one vertical and two horizontal strokes), which belongs to the previous section, we describe the overall similarity below by the (relative unbalanced) minimal total distance of ink transport plus the amount of ink that has to be added or erased between the kanji, which for these examples seems to work rather well. In particular, this is a much more flexible approach than using the label of the components provided in the kanjiVG data or in KRADFILE. Based on the last two, we can only decide whether two components are exactly the same or not. Even that decision is somewhat problematic, because some components do not have labels at all or at least not quite the correct ones (as the corresponding components do not have precise equivalents in Unicode). But much more limiting is the fact that we do not recognize (very) close components in this way, such as the classic radicals \begin{uCJK}⻖\end{uCJK} (Unicode codepoint 2ECF on the left-hand side of kanji as a variant of \begin{uCJK}⾩\end{uCJK}) and \begin{uCJK}⻏\end{uCJK} (Unicode codepoint 2ED6 on the right-hand side of kanji as a variant of \begin{uCJK}⾢\end{uCJK}) or, less extremely, any two consecutive elements in the series \begin{uCJK}牛\end{uCJK}, \begin{uCJK}午\end{uCJK}, \begin{uCJK}干\end{uCJK}, \begin{uCJK}千\end{uCJK} from above. 

Using optimal ink transport raises the question of which font to use. While the top stroke of \begin{uCJK}干\end{uCJK} and \begin{uCJK}千\end{uCJK} may look very similar in certain Japanese gothic fonts (even stroke widths and no decorations), it looks more different in Minch\=o fonts (uneven stroke widths and decorations), such as the one used here, and may look very different in some fonts that depict actual brush strokes. Whether it is better to use a single ``representative'' font or to take an average of total ink transport under a wide selection of fonts is not entirely clear and may depend on the precise target group and goal of the kanji distance. For simplicity, we use the simple stroke path information provided in the kanjiVG data. As it is derived from a schoolbook font, it is rather representative as far as pure path information goes. However, it is surely not representative of the variable stroke width information in various minch\=o and brushstroke fonts. 

Another important point to consider is whether to take any additional knowledge learned with the kanji into account. Even if the top strokes of the  components \begin{uCJK}干\end{uCJK} and \begin{uCJK}千\end{uCJK} look quite similar in a gothic font, a person also learning to write the kanji by hand (and many intermediate to advanced learners in general) can distinguish them quite easily, because they know that these are indeed fundamentally different strokes: a horizontal left-to-right \emph{yoko} in \begin{uCJK}干\end{uCJK}
versus a top-right-to-bottom-left \emph{hidari-harai} in \begin{uCJK}千\end{uCJK}.

Arguably, even more crucial knowledge concerns the meanings and readings of components. Consider \begin{uCJK}去\end{uCJK} and \begin{uCJK}先\end{uCJK}, either as whole-kanji components or as parts of \begin{uCJK}法\end{uCJK} and \begin{uCJK}洗\end{uCJK} for instance. The components may already look somewhat similar on a purely phenomenological level, but the perceived similarity (and the risk of confusing them) increases considerably by the knowledge that they stand for the closely related temporal concepts of \emph{past} and \emph{before}, respectively.

All of these add-ons (typeface, varying stroke widths, stroke types, readings, and meanings) are \emph{not} included in our function $\varrho$ below, although they could be with some individual but overall rather similar problems as the ones we face below.

What we do include, in addition to optimal ink transport, are simple descriptors of position, size, and shape differences. Consider the following ten kanji, which all include a \begin{uCJK}山\end{uCJK} (mountain) component: \begin{uCJK}崔\end{uCJK}, \begin{uCJK}密\end{uCJK}, \begin{uCJK}嵐\end{uCJK}, \begin{uCJK}岬\end{uCJK}, \begin{uCJK}岳\end{uCJK}, \begin{uCJK}崎\end{uCJK}, \begin{uCJK}懲\end{uCJK}, \begin{uCJK}仙\end{uCJK}, \begin{uCJK}岸\end{uCJK},\begin{uCJK}両\end{uCJK}. How would you group them based on similarity? If you have identified the mountain component in all of the kanji (which may take some time, depending on your previous experience), you probably arrive essentially at the three groups
$G_1 = \{\text{\begin{uCJK}岬\end{uCJK}}, \text{\begin{uCJK}崎\end{uCJK}}\}$, $G_2 = \{\text{\begin{uCJK}密\end{uCJK}}, \text{\begin{uCJK}岳\end{uCJK}}\}$, and $G_3=\{\text{\begin{uCJK}崔\end{uCJK}}, \text{\begin{uCJK}嵐\end{uCJK}}, \text{\begin{uCJK}岸\end{uCJK}}\}$. Maybe you did not know what to do with the remaining kanji and placed them all in separate groups; or you assigned, e.g., \begin{uCJK}仙\end{uCJK} to $G_1$, \begin{uCJK}懲\end{uCJK} to $G_3$ and maybe very loosely \begin{uCJK}両\end{uCJK} to $G_2$. The point is that in setting up these three basic groups, you were led by the positions of the mountain components and probably to a lesser extent by their sizes and aspect ratios, all of which agree more or less exactly within each group. You might also find (like me) that $G_1$ differs more from the other two groups than those groups from each other due to the very different aspect ratios of the mountain component. However, this impression depends strongly on how much you see the mountain component as a drawing as opposed to seeing it purely as a component that means mountain.

We note that the mountain component in \begin{uCJK}懲\end{uCJK} is similarly placed and has about the same aspect ratio as the mountain components in~$G_3$. Nevertheless, we perceive \begin{uCJK}懲\end{uCJK} to be quite different from the kanji in $G_3$, which is partly due to the different sizes of the mountain components, but partly also due to a different \emph{exposure} of the component within the kanji. This is another aspect we currently do \emph{not} include in our $\varrho$ below. In $G_3$, the mountains are very exposed, whereas in \begin{uCJK}懲\end{uCJK}, the mountain sits tightly
between three other components.

Apart from kanji with many distinctive components, a lack of exposure is frequently seen due to a single (partly) enclosing component. Compare, e.g., the enclosed components in \begin{uCJK}両\end{uCJK}, \begin{uCJK}風\end{uCJK}, \begin{uCJK}国\end{uCJK}, and \begin{uCJK}園\end{uCJK} with the corresponding (more exposed) components in \begin{uCJK}岳/仙\end{uCJK}, \begin{uCJK}蛍\end{uCJK}, \begin{uCJK}玉\end{uCJK}, and \begin{uCJK}遠\end{uCJK}. The distance we obtain as ink transport plus penalties for position, size, and shape differences may be considerably smaller than the perceived dissimilarity because it ignores the different exposure. This, however, is difficult to judge without data and may be a phenomenon that strongly depends on the individual.

A more general concept than the exposure of a component might be its interaction with neighboring components, which again, is currently not considered in our distance map. It may well be that having an exact match of all components and the same overall structure of the kanji, such as in \begin{uCJK}部\end{uCJK} versus \begin{uCJK}陪\end{uCJK} or \begin{uCJK}招\end{uCJK} and \begin{uCJK}拐\end{uCJK}, creates a stronger sense of similarity than a reasonable additive positional penalty can reflect. This may also extend to cases where the overall structure is not the same, such as for \begin{uCJK}変\end{uCJK} versus \begin{uCJK}赦\end{uCJK}.

\subsection{Optimal transport based comparison of components}  \label{sec:seclevel_wasser}

After sorting out many aspects that we currently do not consider, we now define the remaining quantities, namely the (relative unbalanced) minimal ink transport between standardized components and the penalties for translations, scalings, and distortions.

\begin{figure}[ht]
\begin{center}
  \hspace*{-5mm}\includegraphics[height=18em]{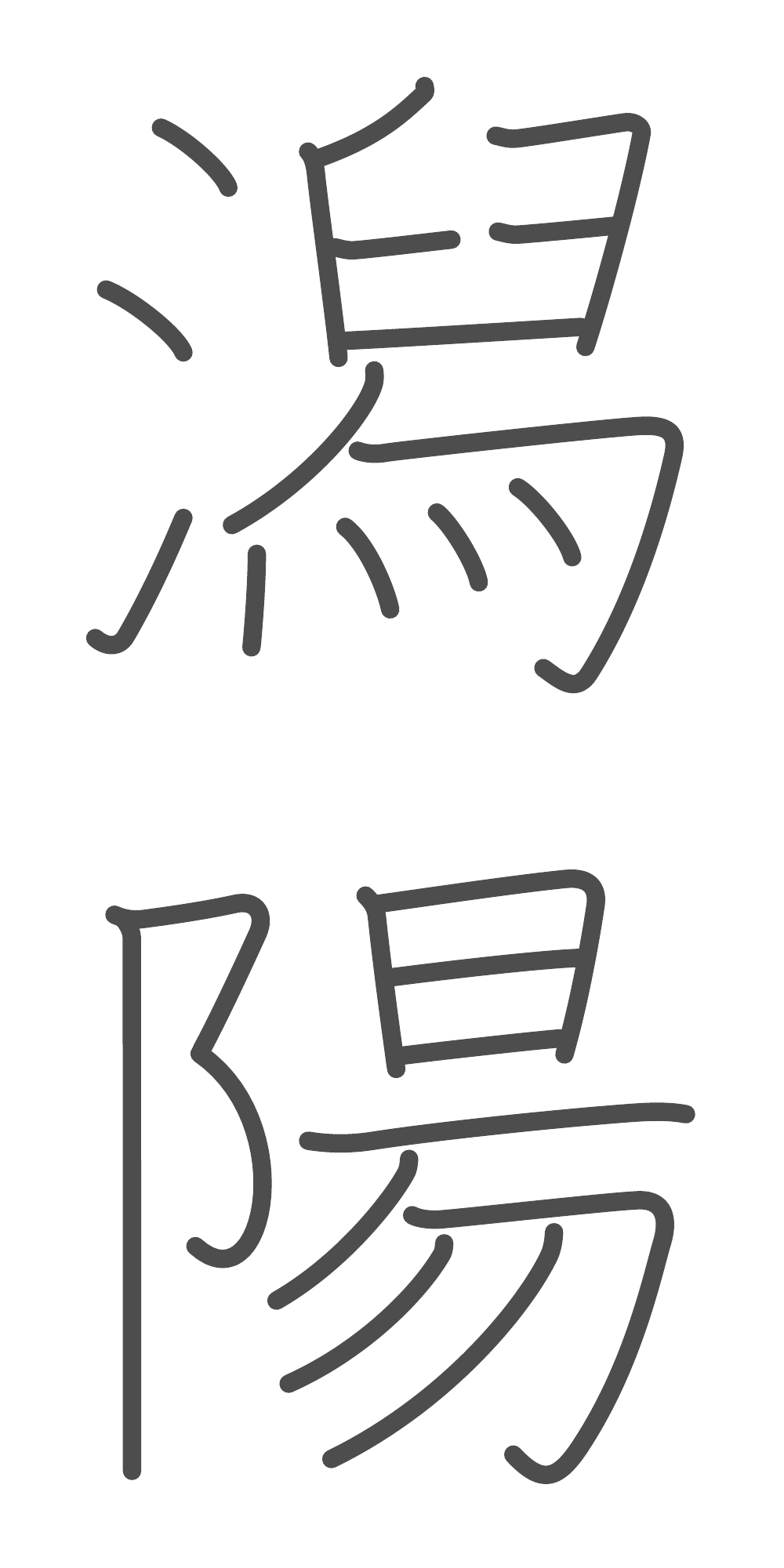}\hspace*{8mm}\includegraphics[height=18em]{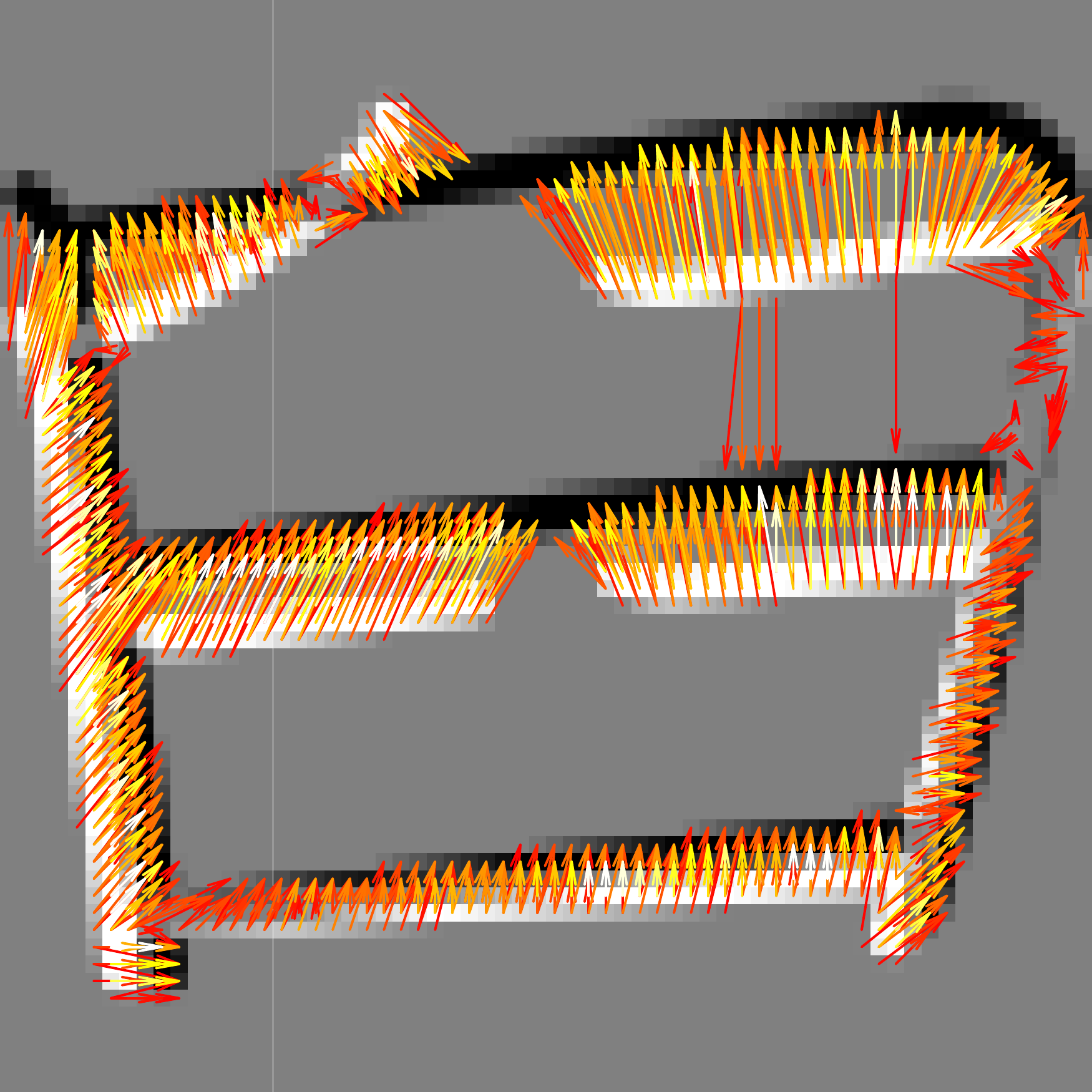}
\end{center}
\vspace*{-4mm}

\caption{\label{fig:brokensun} Unbalanced optimal transport between the top right components of \begin{uCJK}潟\end{uCJK} and \begin{uCJK}陽\end{uCJK}. Left-hand side: the kanji drawn according to the stroke information in kanjiVG. Right-hand side: the grayscale image shows the difference between the components after scaling the maximum of their $x$- and $y$-extensions to $1$. White stands for highly positive values, black for highly negative values; the gray background corresponds to value zero. The arrows give the optimal transport plan between pixels in heat colors (red to orange: smaller amounts of ink, yellow to white: larger amounts of ink). Mass that stays in place is not indicated as it contributes nothing to the transport cost. Parameters are $p=1$, $b=0.4$. We obtain $\raisebox{1.5pt}[0pt][6.25pt]{$\frac{1}{\max\{\abs{\compo}, \abs{\compo'}\}}$} d_{\text{UBW}}(\compo,\compo') = 0.061272$; the translation, log-scale, and log-distortion penalties are 0.041694, 0.363346 and 0.013505, respectively.}
\end{figure}

For the unbalanced minimal ink transport between components $\compo$ and~$\compo'$, we plot them based on the kanjiVG stroke information after aligning their position, size, and distortion. For this, we consider the bounding boxes of the components, which we first center within our canvas $U=[0,1]^2$ and then rescale such that the maximum of their $x$- and $y$-axis is just below~1.\footnote{\label{fn:regi}We currently use bounding boxes for simplicity, but there is considerable room for improvement here. Rather than taking strict bounding boxes, some cropping of outliers could be allowed, e.g., by replacing the minimum and maximum of the ink distribution along the $x$- and $y$-axis by the 5- and 95-percent quantiles. This would improve the match in Figure~\ref{fig:brokensun}, for instance, as it would remove more pixels at the top of the white component than at the top of the black one. There are also more sophisticated image registration methods, which make the components as similar as possible based on a set of transformations.}
We thus obtain two (possibly coarse) pixel images $C=(c_{rs})_{(r,s) \in G}$, $C'=(c'_{rs})_{(r,s) \in G}$, where $G = \{\tfrac{1}{2N}, \tfrac{3}{2N}, \ldots, \tfrac{2N-1}{2N}\}$ is the $N \times N$ grid of pixel centers. Denote by $\delta_{(r,s),(r',s')}$ the Euclidean distance between $(r,s), (r',s') \in G$. We then compute an unbalanced Wasserstein metric between the components (see~\citealp{hkm2023}, \citealp{msm2023} and the references therein) of the form 
\begin{equation} \label{eq:dub}
\begin{split}
  d_{\text{UBW}}(\compo,\compo') = \biggl( \min_{\Pi} \biggl( \sum_{r,s,r',s'} \delta^p_{(r,s), (r',s')} \: \pi_{(r,s), (r',s')} +
        \bigl( &\norm{C}_1 + \norm{C'}_1 - 2\norm{\Pi}_1 \bigr) \frac{b^p}{2} \biggr) \biggr)^{1/p} \\
        &\hspace*{8mm} \in \bigl[0,b\max\bigl\{\norm{C}_1, \norm{C'}_1\bigr\}\bigr], 
\end{split}
\end{equation}
where the minimum is taken over all $\Pi = (\pi_{(r,s), (r',s')}) \in \RR_{+}^{G \times G}$ such that
\begin{equation*}
\begin{split}
  \sum_{r',s'} \pi_{(r,s), (r',s')} &\leq c_{r,s} \quad \text{for all $(r,s) \in G$},\\[-0.5mm]
  \sum_{r,s} \pi_{(r,s), (r',s')} &\leq c'_{r',s'} \quad \text{for all $(r',s') \in G$},\\[-0.5mm]
\end{split}
\end{equation*}
and furthermore
\begin{equation*}
  \norm{C}_1 = \sum_{r,s} c_{r,s}, \ \norm{C'}_1 = \sum_{r',s'} c'_{r',s'}, \text{ } \norm{\Pi}_1 = \sum_{r,s,r',s'} \pi_{(r,s),(r',s')}.
\end{equation*}
In other words, $d_{\text{UBW}}(\compo,\compo')$ gives the cost of transforming normalized versions of $\compo$ into $\compo'$ in terms of minimal total ink transport, allowing the option of adding/deleting ink at the cost of $b/2^{1/p}$ per unit.\footnote{Dividing by $2$ for the factor $\frac{b^p}{2}$ in the above formula is debatable. Here we do it to ensure that the factor $2$ does not show up in the upper bound of the distance. On the other hand, if we omit the division by $2$, we have more intuitive costs of $b$ per unit for ink added/deleted, and the $d_{\text{UBW}}$-distance still only very rarely goes beyond $b\max\bigl\{\norm{C}_1, \norm{C'}_1\bigr\}$ for actual components.} We always choose $p=1$ and $b=0.4$ (i.e., 0.4 is the maximal distance at which we assume that ink is associated, corresponding to a cost of 0.2 for adding and 0.2 for deleting the ink).  
Figure~\ref{fig:brokensun} gives an example for the top right components of \begin{uCJK}潟\end{uCJK} and \begin{uCJK}陽\end{uCJK}.

It is tempting to compute $d_{\text{UBW}}(\compo,\compo')$ between the raw components instead of normalizing them first and adding penalties for translation, scaling, and distortion afterwards. However, in doing so, we would undervalue the strong effect of shape similarity. For example, \begin{uCJK}部\end{uCJK} and \begin{uCJK}陪\end{uCJK} could never have a (reasonably) small overall $d$-distance because the $d_{\text{UBW}}$-distance between the similar-looking components would be close to the maximal possible value since they are shifted by just about $0.4=b$. Even the $d_{\text{UBW}}$-distance between the very different components at the same position is smaller than that.

Below we always consider \emph{relative} $d_{\text{UBW}}$-distances, i.e., we divide by $b \max\{\norm{X}_1, \norm{Y}_1\}$,
which brings the maximum distance to $1$. While $d_{\text{UBW}}$ is always a metric, division by $b \max\{\norm{X}_1, \norm{Y}_1\}$ destroys the triangle inequality unless we only consider kanji whose total amount of ink is within a certain range (\citealp{SandWirth2023}). However, the relative $d_{\text{UBW}}$-distance map agrees much better with the way how we perceive dissimilarities: Adding a single stroke to a component that already has many strokes makes for a smaller dissimilarity than adding it to a component with few strokes, compare e.g.\ \begin{uCJK}史\end{uCJK} and \begin{uCJK}吏\end{uCJK} as opposed to \begin{uCJK}丨\end{uCJK} and \begin{uCJK}十\end{uCJK}.

We next describe the translation, scaling, and distortion penalties. Consistently with the normalization of components above, we base these penalties on the bounding boxes of the components. The translation penalty $\tau(\compo, \compo')$ is simply the Euclidean distance between the centers of the bounding boxes. We use the geometric mean of the side lengths of the bounding box as scale and the ratio of these side lengths as distortion. We then define the scale penalty $\sigma(\compo, \compo')$ as the absolute log-value of the scale ratio and the distortion penalty $\chi(\compo, \compo')$ as the absolute log-value of the distortion ratio between the two components. 

For now, we take a linear combination of increasing functions of the relative $d_{\text{UBW}}$-distance and the various penalty terms. More sophisticated choices are discussed in Subsection~\ref{ssec:learning}. Applying an increasing function appears essential as our perception does not seem to be linear in these terms at all: If the relative $d_{\text{UBW}}$-distance, say, between components is very small, they are easily perceived as similar. If it is somewhat small, the similarity may not be consciously perceived. Still, we might confuse the components at some later point when memory becomes blurred (see e.g.\ the right parts of the kanji in Figure~\ref{fig:brokensun}, which we would probably not have confused with each other at first glance). Once a certain distance level is crossed, it is doubtful that our brains ever associate the components with each other. For our three penalties, we suspect a similar behavior, albeit with more interplay with other effects.

\begin{figure}
\begin{center}
  \hspace*{-3mm}\includegraphics[width=0.45\textwidth]{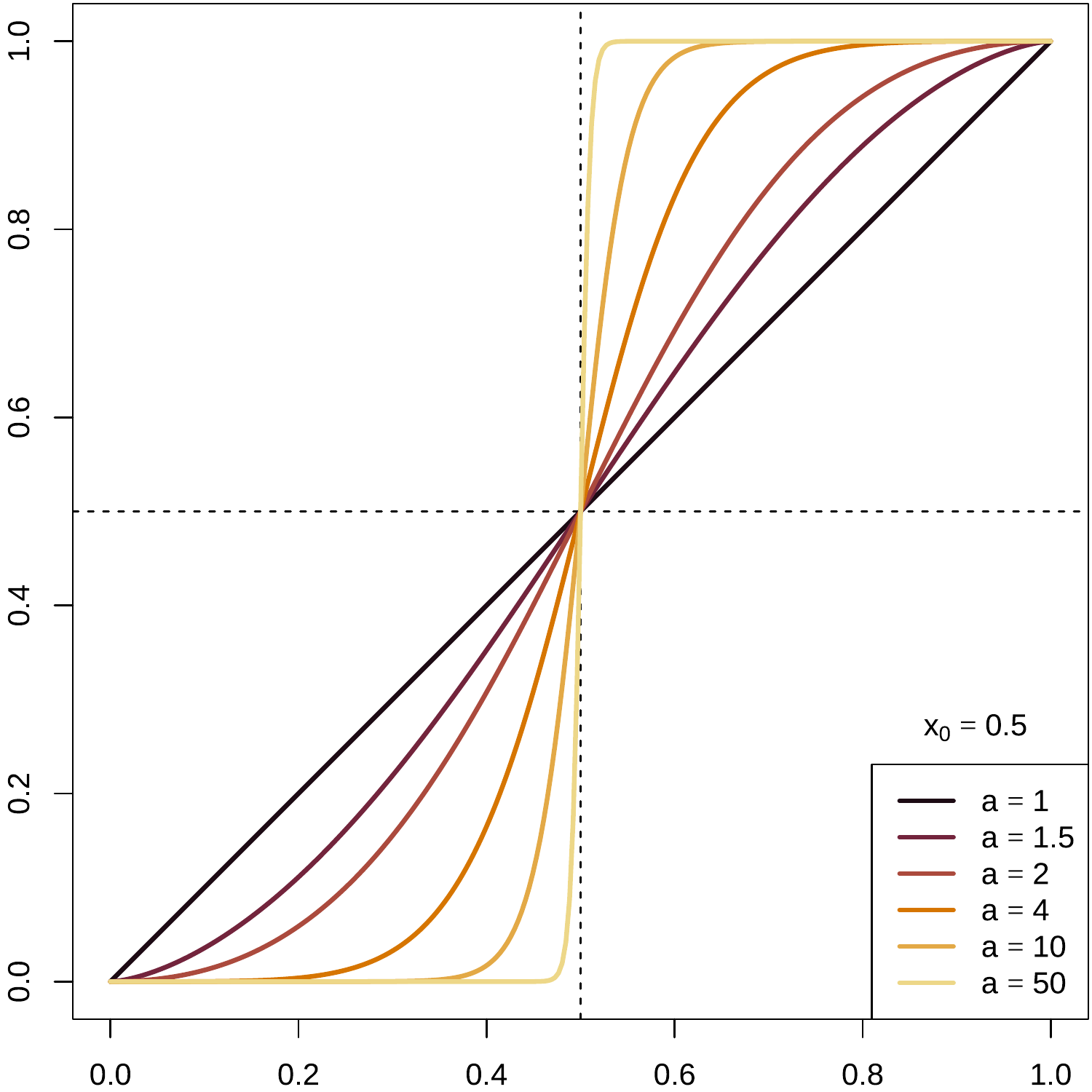}\hspace*{8mm}\includegraphics[width=0.45\textwidth]{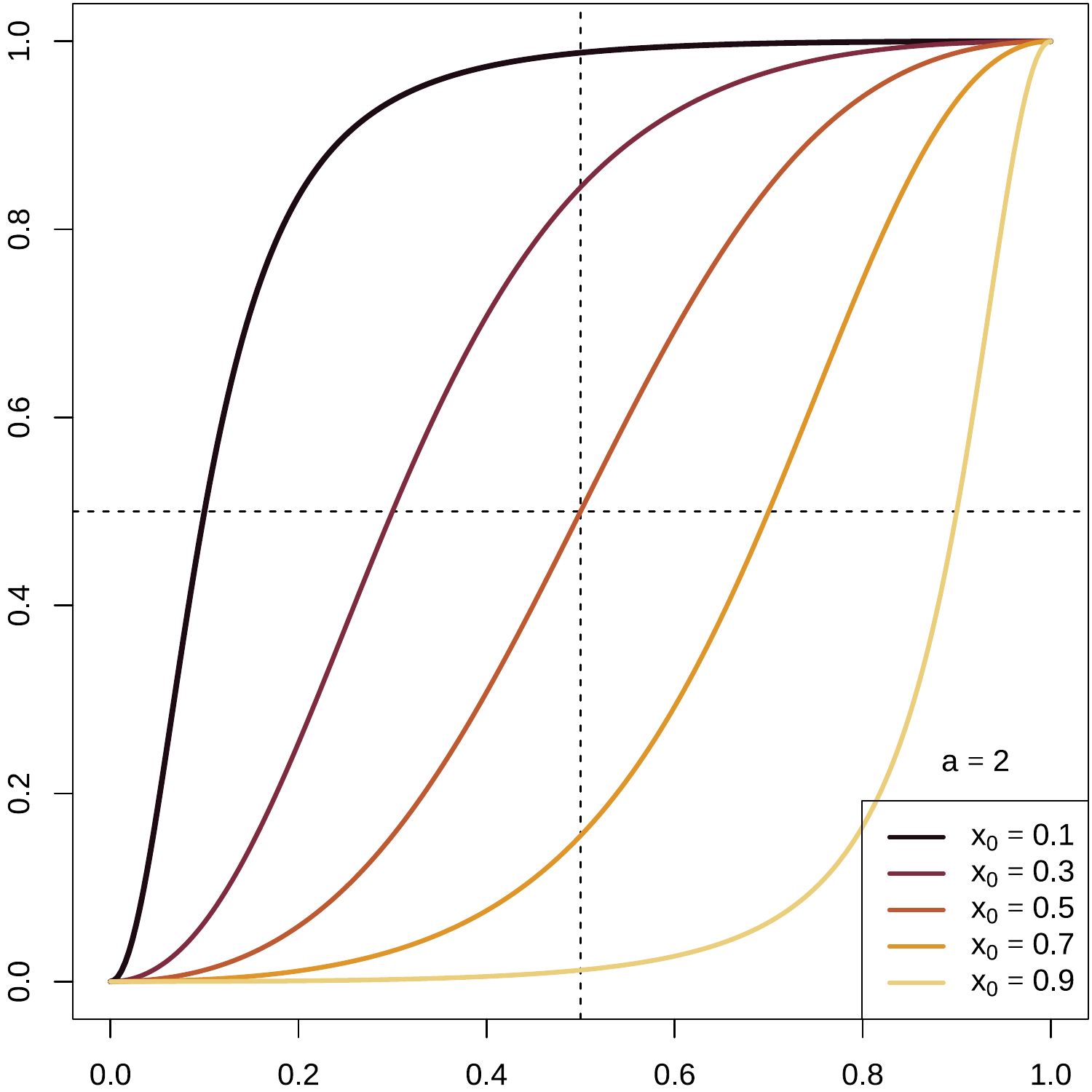}
\end{center}
\vspace*{-5mm}

\caption{\label{fig:logi} Functions $\psi$ for transforming distances and transformation, scaling, and distortion penalties. The parameter $a$ determines the shape, interpolating between the identity on $[0,1]$ ($a=1$) and the step function $1_{[1/2,1]}$ ($a \to \infty$). The parameter $x_0$ determines the point $x$ where $\psi(x)=1/2$.}
\end{figure}

We apply then increasing functions $\psi_{\text{dist}}$, $\psi_{\text{trans}}$, $\psi_{\text{scale}}$, $\psi_{\text{distort}}$ $\colon [0,1] \to [0,1]$ to the respective quantities that tone down small values and blow up larger ones. For the sake of simplicity, we choose
\begin{equation*}
  \psi(x) = \psi^{(\alpha,x_0)}(x) = \frac{1}{1+\bigl( \frac{x_0}{1-x_0} \frac{1-x}{x} \bigr)^\alpha}, \quad x \in [0,1],
\end{equation*}
for parameters $\alpha \geq 1$ and $x_0 \in (0,1)$ determining the shape and shift of the transform, respectively, see Figure~\ref{fig:logi}. This function is obtained by plugging a standard logit into a general logistic function and reparametrizing the shift.\footnote{The standard logit function is $(0,1) \to \RR, u \mapsto \log(\frac{u}{1-u})$, the general logistic function $\RR \to (0,1), y \mapsto \bigl(1+\exp(-\alpha(y - \beta))\bigr)^{-1}$ for parameters $\alpha>0$ (here) and $\beta \in \RR$.} 

So finally, combining all the discussed ingredients, we obtain for $\varrho$
\begin{equation}  \label{eq:rhoprop}
\begin{split}
  \varrho(\compo, \compo') &= \lambda_0 \, \psi_0 \bigl(\raisebox{1pt}{$\tfrac{1}{b \max\{\abs{\compo}, \abs{\compo'}\}}$} \hbit d_{\text{UBW}}(\compo,\compo')\bigr) \\[1mm]
           & \hspace*{6mm}   + \lambda_1 \, \psi_1 \bigl(\tau(\compo,\compo')\bigr)
                             + \lambda_2 \, \psi_2 \bigl(\sigma(\compo,\compo')\bigr)
                             + \lambda_3 \, \psi_3 \bigl(\chi(\compo,\compo')\bigr),
\end{split}
\end{equation}
where $\lambda_i \geq 0$ are parameters satisfying $\sum_{i=0}^3 \lambda_i = 1$. In the absence of data and a specific target for our overall distance, we resort to some educated trial and error for choosing the various parameters in Section~\ref{sec:results}; see~Equation~\eqref{eq:concrete_rho} for the concrete parameter choices. As soon as (a sufficient amount of) data is available, it seems desirable to include also interaction terms in the linear expression on the right-hand side; see also the next subsection.

\subsection{Estimating the function $\bs{\varrho}$ from data} \label{ssec:learning}

In the long run, it is desirable to estimate $\varrho(\compo, \compo')$ as a function of $\frac{1}{\max\{\abs{\compo}, \abs{\compo'}\}} d_{\text{UBW}}(\compo,\compo')$, $\tau(\compo,\compo')$, $\sigma(\compo,\compo')$ and $\chi(\compo,\compo')$ (or other suitable penalties for the transformations used to register the components) from data. Test items for collecting such data could be similar to those already available in the literature; see \cite{YenckenBaldwin2006, YenckenBaldwin2008}. Examples are (presenting two kanji) ``Rate the similarity on a continuous scale from 0 to 1'' or  (presenting three kanji) ``Is kanji A or kanji B closer to kanji C?''. Note, however, that these items are used here in the first place for training rather than testing the kanji distance map, which means that they can (and should) be designed strategically. 

Two very different scenarios are possible. On the one hand, we can collect a small amount of data from an individual learner or a small homogeneous group with a few targeted questions. These could have the specific goal of first learning the parameters $\psi_i$, $i=0,1,2,3$, then their factors $\lambda_i$. E.g., estimation of $\psi_0$ would be based on presenting a number of individual components and simple kanji without emphatic decompositions, such as any two from the set $\{\text{\begin{uCJK}牛\end{uCJK}}, \text{\begin{uCJK}午\end{uCJK}}, \text{\begin{uCJK}干\end{uCJK}}, \text{\begin{uCJK}千\end{uCJK}}\}$, but also more obviously different kanji, like (pairs from) $\{\text{\begin{uCJK}日\end{uCJK}}, \text{\begin{uCJK}月\end{uCJK}}, \text{\begin{uCJK}目\end{uCJK}}\}$, $\{\text{\begin{uCJK}山\end{uCJK}}, \text{\begin{uCJK}川\end{uCJK}}\}$, $\{\text{\begin{uCJK}老\end{uCJK}}, \text{\begin{uCJK}古\end{uCJK}}\}$ or $\{\text{\begin{uCJK}女\end{uCJK}}, \text{\begin{uCJK}名\end{uCJK}}\}$. The functions $\psi_1, \psi_2$ and $\psi_3$ would be estimated based on more complex kanji with clear components, between which one identifiable component moves around and gets scaled and distorted by various amounts, whereas other components are not readily associated with one another. Examples include $\{\text{\begin{uCJK}案\end{uCJK}}, \text{\begin{uCJK}査\end{uCJK}}\}$, $\{\text{\begin{uCJK}賀\end{uCJK}}, \text{\begin{uCJK}程\end{uCJK}}\}$ or $\{\text{\begin{uCJK}喋\end{uCJK}}, \text{\begin{uCJK}割\end{uCJK}}, \text{\begin{uCJK}鉛\end{uCJK}}\}$. 
All of these examples were generated ad hoc, but to increase efficiency, the kanji sets presented could well be designed from the actual values of $\raisebox{1.5pt}{$\frac{1}{\max\{\abs{\compo}, \abs{\compo'}\}}$} d_{\text{UBW}}(\compo,\compo')$, $\tau(\compo,\compo')$, $\sigma(\compo,\compo')$ and $\chi(\compo,\compo')$.

The estimators for the $\psi_i$ and the $\lambda_i$ are most easily based on linear regression. For $\psi_i$, we can first apply a logit transform to both the dependent and independent variable (the perceived similarity on a scale of 0 to 1 and the actual $\tau(\compo,\compo')$-value, say) and work with the transformed parameter $\tilde{x}_0 = \alpha \log(\frac{1-x_0}{x_0})$. 

On the other hand, when a large amount of data is available from questions asked to a sizeable group of heterogeneous learners, a more complex function $\varrho$ of optimal ink transport and registration penalty than \eqref{eq:rhoprop} can be estimated. Such a function may contain further covariates (e.g., \ from the specific learner) and has either many parameters based on a flexible design or is non-parametric. Denoting the (possibly extended) generic covariate vector by $x$ and the (training) data by $\xi = (x_i)_{1 \leq i \leq n}$ and $\eta = (y_i)_{1 \leq i \leq n}$ (perceived similarities), suitable estimation techniques include kernel smoothing methods (see~\citealp{HastieEtAl2009}), such as the (transformed) Nadaraya--Watson estimator
\begin{equation*}
    \hat{\varrho}(x; (\xi, \eta)) = \psi \Biggl( \frac{\sum_{i=1}^n y_i \, k\bigl(\tfrac{x-x_i}{h})}{\sum_{i=1}^n k\bigl(\tfrac{x-x_i}{h})} \Biggr),
\end{equation*}
where $k$ is a suitable kernel on the input space, $h > 0$ a bandwidth parameter,\footnote{More generally, replace $k\bigl(\tfrac{x-x_i}{h}\bigr)$ by $k\bigl(H^{-1/2}(x-x_i)\bigr)$, where $H$ is a positive definite bandwidth matrix.} and $\psi \colon \RR \to [0,1]$ a logistic function. 

Alternatively, if we care less about explanatory features of the estimator, neural networks may be used, such as a simple feedforward network of the form
\begin{equation*}
    \hat{\varrho}(x; (\xi, \eta)) = f_T \bigl( A_{\hat{\theta}_T(\eta, \xi)} \bigl( f_{T-1}\bigl(A_{\hat{\theta}_{T-1}(\eta, \xi)} \ldots f_1(A_{\hat{\theta}_1(\eta, \xi)}(x))\bigr)\bigr)\bigr),
\end{equation*}
where $A_{\theta_t}$ denote affine maps with parameter vector $\theta_t$ and $f_t$ are prespecified (non-linear) activation function. Here $\hat{\theta}_{t}(\xi, \eta)$ denote the parameter vectors learned by some form of gradient descent algorithm based on the training data.

\section{A concrete kanji distance map and first results} \label{sec:results}

Let us choose the parameters left open in~\eqref{eq:rhoprop} to obtain a first concrete distance map as a proof of concept. We do this in a simple way based on the discussion so far and some trial and error, but without data. As before we set $p=1$ and $b=0.4$ in $d_{UBW}$. For simplicity, we take for $\psi_1$, $\psi_2$ and $\psi_3$ the identity functions. However, we implement two measures to increase the similarity focus of the $\psi_0$-term, incorporating the effect that near-similarity is more strongly perceived than the $d_{\text{UBW}}$-distance term suggests. On the one hand, we simply choose $\alpha=2$ and $x_0=0.4$ for $\psi_0$; on the other hand, we include the component labels from the kanjiVG data and allow them to overrule $d_{\text{UBW}}(\compo,\compo')$ by setting it to zero if labels for both components are present and match. Regarding the $\lambda_i$, it is clear that $\lambda_0$ should take most of the weight by far since there will always be assignments of components with very small translation, scale, and distortion costs, but if these components are not similar in terms of optimal ink transport, the distance should stay large. I have opted for $\lambda_0=0.8$ and split the remaining $\lambda_i$ among the transformation penalties after looking at some concrete examples of how the penalties behave to one another. Overall we use then instead of \eqref{eq:rhoprop}
\begin{equation} \label{eq:concrete_rho}
\begin{split}
  \varrho(\compo, \compo') &= 0.8 \, \psi^{(2,\hbit0.4)} \Bigl(\tfrac{\one\{\ell(\compo) \neq \ell(\compo')\}}{0.2 \max\{\abs{\compo}, \abs{\compo'}\}} \hbit d_{\text{UBW}}(\compo,\compo')\Bigr) \\[1mm]
  &\hspace*{6mm}  + 0.1 \, \tau(\compo,\compo') + 0.05 \, \sigma(\compo,\compo') + 0.05 \, \chi(\compo,\compo').
\end{split}
\end{equation}

The other quantities relevant for the kanji distance in \eqref{eq:kanjidist} are $a=0.25$,\footnote{The penalty for unmatched weight in the overall component comparison. Do not mix up with $\alpha$.}
the weights $w_{l,i}^{(1)}$, $w_{l,i}^{(2)}$ as in \eqref{eq:compoweights} with a trickle loss of 0.02 after the second level, and finally $\mu(w, w') = \min(w, w')$.

Components and their labels stem from the kanjiVG website but underwent further processing. The kanjiVG data consists of SVG files that store components hierarchically in nested \verb|<g>|-elements. There are, however, also many such elements that contain additional information, creating artificial hierarchy levels. Moreover, the files respect the stroke orders, meaning logical components are sometimes split into several parts. To obtain the components used in~\eqref{eq:concrete_rho}, we re-united these parts and removed some of the (for our purposes) unnecessary hierarchies, erring rather on the side of caution. The resulting components are part of the data set \textsf{kvec} that is available in the \textsf{R} package \textsf{kanjistat}. 

We perform several experiments with the distance map in~\eqref{eq:concrete_rho}. First, the nearest neighbors of the four kanji discussed in the introduction were determined (Table~\ref{tab:k90neighbors_kdist}). Because of the close match of \begin{uCJK}粋\end{uCJK} and \begin{uCJK}枠\end{uCJK}, we would like to investigate the neighborhood of these two kanji further. For this, we take the eight nearest neighbors of each of \begin{uCJK}粋\end{uCJK} and \begin{uCJK}枠\end{uCJK}, which turn out to coincide with the kanji at distance $< 0.14$ of these kanji, and determine their four nearest neighbors in turn. This results in 42 unique kanji, between which we compute all $\binom{42}{2} = 861$ pairwise distances.

\begin{figure}
\begin{center}
  \hspace*{-8mm}\includegraphics[width=0.58\textwidth]{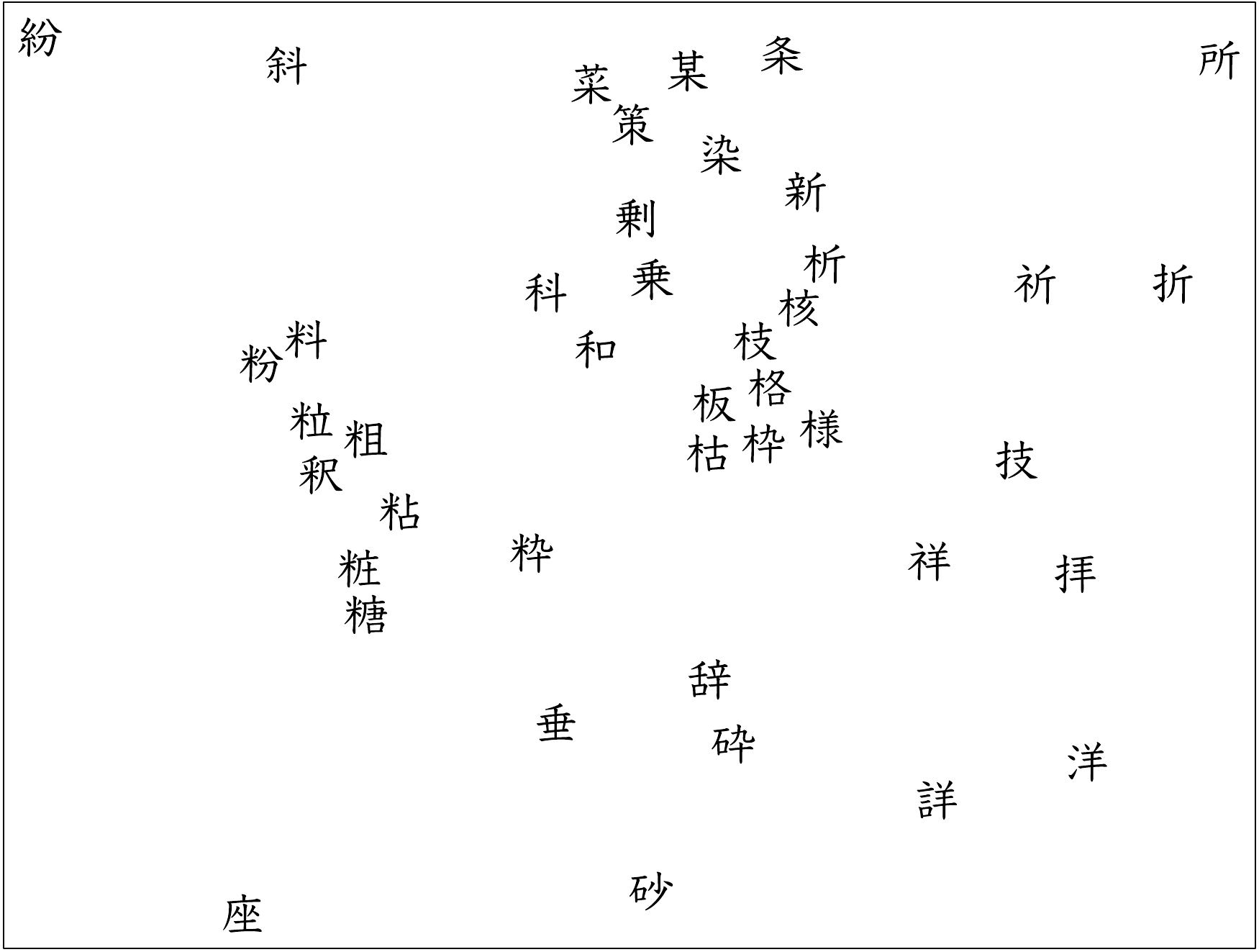}\hspace*{-0.5mm}\includegraphics[width=0.43\textwidth]{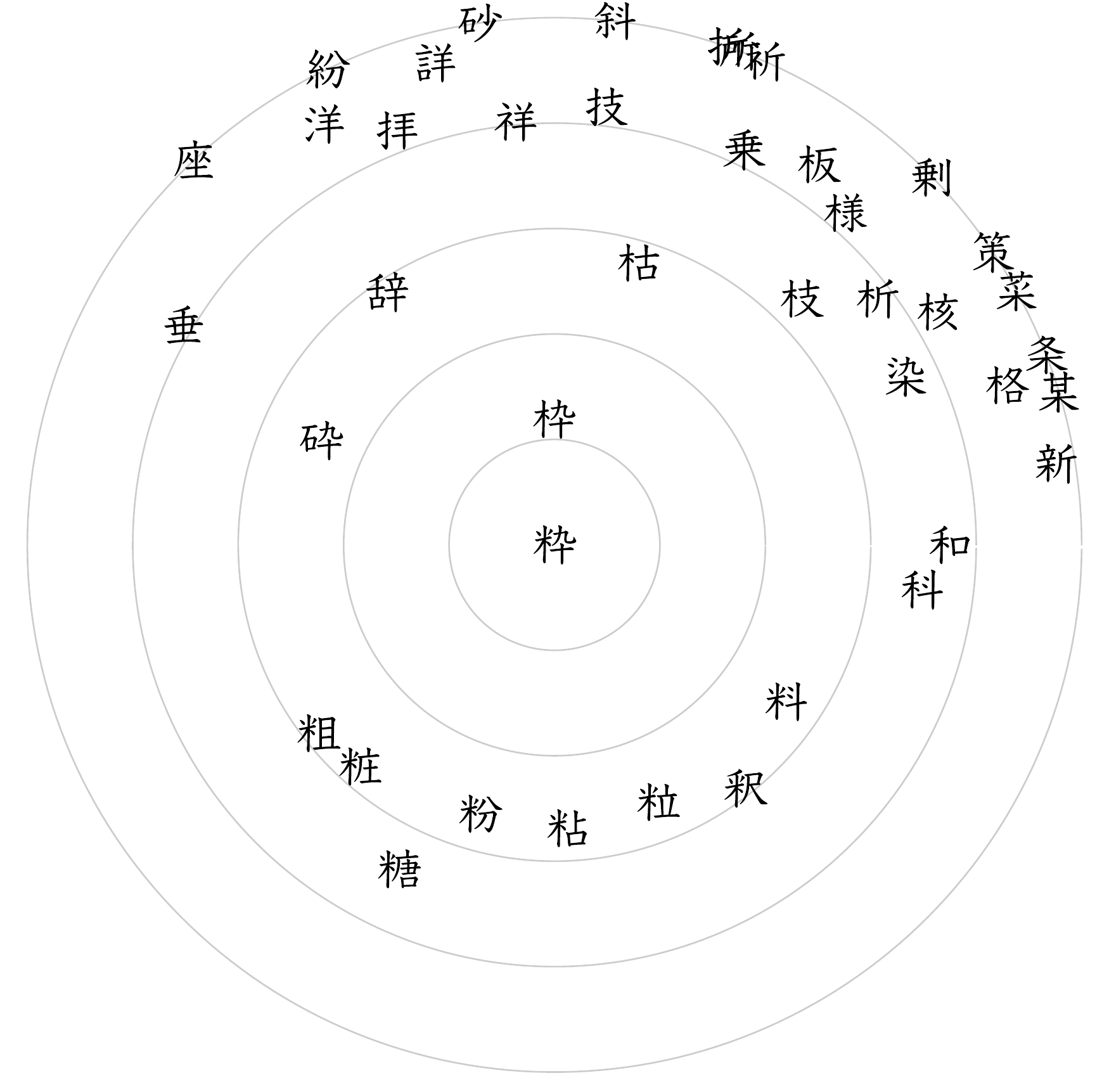}\hspace*{-10mm}
\end{center}
\vspace*{-7mm}

\caption{\label{fig:mapof42} 42 kanji based on neighbors of neighbors of \begin{uCJK}粋\end{uCJK} and \begin{uCJK}枠\end{uCJK}. \textit{Left:} Global map based on non-metric (aka monotone or ordinal) multidimensional scaling. Note that the purpose is depicting the global neighborhood structure, not obtaining good clusters. \textit{Right:} The neighborhood seen from \begin{uCJK}粋\end{uCJK} based on focused (metric) multidimensional scaling. The circles indicate multiples of distance 0.05 from \begin{uCJK}粋\end{uCJK}.}
\end{figure}

In Figure~\ref{fig:mapof42} (left), we see an overall map of these kanji produced by Kruskal's non-metric multidimensional scaling (nMDS) using the \textsf{R} package \textsf{vegan} (\citealp{vegan}). The distances are strongly distorted in several obvious ways (in fact, \begin{uCJK}粋\end{uCJK} and \begin{uCJK}枠\end{uCJK} should be very close, only just beaten by \begin{uCJK}析\end{uCJK} and \begin{uCJK}祈\end{uCJK}, which on the map are not very close either). This happens because only two dimensions are available and because nMDS aims, in the first place, to keep the parity of distance comparisons rather than the actual distances themselves. Nevertheless, the map is useful for spotting similarities among the 42 kanji.

A remarkable feature is the (sometimes loose and overlapping) groups of kanji with the same structure. Note, in particular, the five kanji in the middle on top,
which have the tree component at the bottom and a global top-bottom split rather than a left-right split. This grouping is notable since we did not explicitly model this type of split in our distance map. It gets automatically picked up by the distance map as, among the tree (and possibly other matched) components, the translation, scaling, and distortion penalties are virtually zero.

The right-hand side of Figure~\ref{fig:mapof42} gives a more targeted view of the neighborhood of a single kanji. It is based on a focused (metric) multidimensional scaling, see~\cite{UrpaAnders2019focusedMDS}. This means that we display the exact distance from the central kanji \begin{uCJK}粋\end{uCJK} to every other kanji. The placement of each subsequent kanji on the circle corresponding to this distance is determined by minimizing the stress, i.e., the sum of quadratic deviations, of all distances to the kanji already placed. The \textsf{R} package \textsf{focusMDS} provides a dynamic version of this plot, where the user can click her way through the various kanji. 

As expected, due to some of the very small distances between similar kanji (e.g., between \begin{uCJK}粋\end{uCJK} and \begin{uCJK}枠\end{uCJK}), the triangle inequality is not always satisfied among these 42 kanji. However, from among $\binom{42}{3} = 11,\!480$ triplets of kanji, only 57 triangle inequalities are violated. Most of these violations include a very small distance ($<0.08$) or are very mild. For the few closest kanji, the effects in the dynamic version of focusedMDS are, therefore, a bit strange sometimes, but overall this is still a very helpful representation.

Of course, several more sophisticated methods exist to depict maps based on similarities or distances; see e.g.\ the methods compared in \cite{AmidEtAl2022}.

For a final demonstration, we pick 100 kanji uniformly at random from among the 2136 J\=oy\=o kanji and present their 16 nearest J\=oy\=o neighbors in the appendix. Skimming through the table, the similar kanji and their left-to-right order (including the colors indicating brackets of distances) seem mostly reasonable. Overall these results appear very useful if the goal is to generate a set of similar kanji from which the learner can choose what she finds interesting or what she is looking for.

On the other hand, there are sometimes a few kanji at a pretty early position that would probably not be seen as close by (m)any learners. One example is for \begin{uCJK}悔\end{uCJK} (twelfth line from top), where we have \begin{uCJK}母\end{uCJK} at the sixth position with a distance of 0.123339, which is due to the close match with the bottom right.
Another, probably worse example is for \begin{uCJK}残\end{uCJK} (third line from the bottom), where we have \begin{uCJK}三\end{uCJK} at the fifth position with a distance of 0.1352396. Upon inspection, this distance is so small, partly because the match of \begin{uCJK}三\end{uCJK} with the three horizontal strokes on the right-hand side is allowed at ink transportation cost zero (because the kanjiVG labels match), but mainly because this match gets an inappropriately large weight of 0.6055 due to the fact that strokes 5--10 of \begin{uCJK}残\end{uCJK} (its whole right-hand side) are subsumed under the component \begin{uCJK}三\end{uCJK} in the kanjiVG file for some reason.

\section{Conclusion}

In this article, we have introduced a general framework for a kanji distance map based on matching hierarchical structures of components in a nesting-free way across all levels. The underlying component distance map is typically built on a combination of the relative cost of unbalanced optimal ink transport of registered components and an appropriate cost for the registration. We have discussed many other aspects that might influence the component cost but are currently not included in our first kanji distance~\eqref{eq:rhoprop}.

This first kanji distance is a prototype that gives promising results. 
However, the hard work has only just begun. In order to arrive at an authoritative general-purpose kanji distance map (or also more targeted distance maps for specific audiences), the following steps are required.

\begin{enumerate}
  \item Errors and misinterpretations in the component data sets must be further reduced. In my experience, the kanjiVG data set is the best starting point that is publically available. However, its structure is rather involved due to the many concurrent goals pursued in this data set. Our current code for transforming the kanjiVG files into suitable component structures does not fully treat all the special cases, which sometimes are difficult to understand even for a human interpreter of the files.\\
  In a similar vein, the labels we used for identifying the components, taken from the \verb|element|-attributes in kanjiVG, are neither complete nor unique. It is desirable to fix this, e.g., by creating extended component labels that also include the \verb|type|-attributes of all the strokes involved or by trying to combine the data with decomposition information from other sources, such as KRADFILE.
  \item A more sophisticated registration procedure should be used (see Footnote~\ref{fn:regi} for example) with a penalty term that either evaluates the necessary transformation as a whole or can account for interactions between translation, scaling, and distortion effects.
  \item Experiments should be devised and conducted to obtain suitable data for training and evaluation. An ideal target group may be lower-semester Japanology students with some exposure to kanji due to their current studies but no previous Chinese script background. We want to exclude subjects with no prior exposure to kanji as the heterogeneity of the population might be too large with respect to how similarity differences are perceived. Also, we might exclude subjects at advanced levels of kanji knowledge as for this group, it is harder to control what previous knowledge enters the similarity judgment.
  \item The whole process should be supervised by an expert in cognitive psychology, psycholinguistics, or (ideally) kanji/hanzi learning theory.
\end{enumerate}

Regarding the mention of Chinese script in the last point, we note that many aspects of the distance maps described in this article carry over directly to (simplified and traditional) Chinese hanzi characters. The only major hurdle is that other data sets are required to obtain decompositions, labels, etc.

As a final warning, we note that care should be taken not to over-engineer the kanji distance. The present paper offers a range of tools and ideas that can be combined rather freely. However, the goal should be to make a judicious selection for the situation at hand based on expert knowledge, statistical data analysis, and common sense.

\section*{Appendix: the night parade of 100 kanji}

To allow the reader to evaluate to what extent the kanji distance with the concrete parameter choices from Chapter~\ref{sec:results} is in accordance with her own judgment, we have computed nearest neighbors within the set of J\=oy\=o kanji for 100 randomly sampled characters. For objectivity and reproducibility, the kanji have been sampled with the command \verb|sample(2136, 100)| (referring to the indices in the \verb|kbase|, \verb|kmorph|, and \verb|kvec| datasets in \textsf{kanjistat}) directly at the startup of \textsf{R} version 4.2.2.\footnote{The starting seed should be stable across (almost) all systems and widely stable across different versions. The first five indices were 360, 555, 183, 1776, 1220.} 

The following table gives the 16 nearest neighbors for each of the kanji in the order of their distances, where we use different colors ranging from darker red (small distance) to white (large distance) for the background of the kanji as follows:
\definecolor{peach1}{HTML}{EA4C3B}
\definecolor{peach2}{HTML}{F0724B}
\definecolor{peach3}{HTML}{F49265}
\definecolor{peach4}{HTML}{F7AE83}
\definecolor{peach5}{HTML}{F9C8A4}
\definecolor{peach6}{HTML}{FADDC3}
\begin{equation*}
 0 \textcolor{peach1}{-\blacksquare-} 0.075 \textcolor{peach2}{-\blacksquare-} 0.1 \textcolor{peach3}{-\blacksquare-} 0.125
   \textcolor{peach4}{-\blacksquare-} 0.15  \textcolor{peach5}{-\blacksquare-} 0.175 \textcolor{peach6}{-\blacksquare-} 0.2
\end{equation*}

\begin{longtable}{|c||c|c|c|c|c|c|c|c|c|c|c|c|c|c|c|c|}
  \hline
  \hline
\begin{uCJK}速\end{uCJK} & \cellcolor{peach1}\begin{uCJK}述\end{uCJK} & \cellcolor{peach2}\begin{uCJK}連\end{uCJK} & \cellcolor{peach2}\begin{uCJK}遠\end{uCJK} & \cellcolor{peach2}\begin{uCJK}疎\end{uCJK} & \cellcolor{peach3}\begin{uCJK}達\end{uCJK} & \cellcolor{peach3}\begin{uCJK}辣\end{uCJK} & \cellcolor{peach3}\begin{uCJK}束\end{uCJK} & \cellcolor{peach3}\begin{uCJK}造\end{uCJK} & \cellcolor{peach3}\begin{uCJK}適\end{uCJK} & \cellcolor{peach3}\begin{uCJK}味\end{uCJK} & \cellcolor{peach3}\begin{uCJK}保\end{uCJK} & \cellcolor{peach3}\begin{uCJK}和\end{uCJK} & \cellcolor{peach3}\begin{uCJK}違\end{uCJK} & \cellcolor{peach4}\begin{uCJK}逓\end{uCJK} & \cellcolor{peach4}\begin{uCJK}逮\end{uCJK} & \cellcolor{peach4}\begin{uCJK}架\end{uCJK} \\ 
   \hline
\begin{uCJK}敗\end{uCJK} & \cellcolor{peach2}\begin{uCJK}散\end{uCJK} & \cellcolor{peach3}\begin{uCJK}致\end{uCJK} & \cellcolor{peach3}\begin{uCJK}敢\end{uCJK} & \cellcolor{peach3}\begin{uCJK}貯\end{uCJK} & \cellcolor{peach3}\begin{uCJK}貼\end{uCJK} & \cellcolor{peach3}\begin{uCJK}財\end{uCJK} & \cellcolor{peach3}\begin{uCJK}則\end{uCJK} & \cellcolor{peach4}\begin{uCJK}測\end{uCJK} & \cellcolor{peach4}\begin{uCJK}欺\end{uCJK} & \cellcolor{peach4}\begin{uCJK}放\end{uCJK} & \cellcolor{peach4}\begin{uCJK}政\end{uCJK} & \cellcolor{peach4}\begin{uCJK}枚\end{uCJK} & \cellcolor{peach4}\begin{uCJK}数\end{uCJK} & \cellcolor{peach4}\begin{uCJK}牧\end{uCJK} & \cellcolor{peach4}\begin{uCJK}救\end{uCJK} & \cellcolor{peach4}\begin{uCJK}賠\end{uCJK} \\
\hline
\hline
\end{longtable}


\end{document}